\documentclass[fleqn,usenatbib]{mnras}

\usepackage{newtxtext,newtxmath}

\usepackage[T1]{fontenc}

\DeclareRobustCommand{\VAN}[3]{#2}
\let\VANthebibliography\thebibliography
\def\thebibliography{\DeclareRobustCommand{\VAN}[3]{##3}\VANthebibliography}

\usepackage{graphicx}	
\usepackage{amsmath}	
\DeclareMathAlphabet{\pazocal}{OMS}{zplm}{m}{n}
\usepackage{afterpage}
\usepackage{lscape}
\usepackage{rotating}
\usepackage{longtable}
\usepackage{multicol}
\newsavebox\ltmcbox

\title[BEBOP-3 circumbinary detection]{BEBOP VII. SOPHIE discovery of BEBOP-3b, a circumbinary giant planet on an eccentric orbit}

\author[T A. Baycroft et al.]{
Thomas A. Baycroft,$^{1}$\thanks{E-mail: t.baycroft@bham.ac.uk - thomasbaycroftastro@gmail.com}
Alexandre Santerne,$^{2}$
Amaury H.M.J. Triaud,$^{1}$
Neda Heidari,$^{3}$
Daniel Sebastian,$^{1}$
\newauthor
Yasmin T. Davis,$^{1}$
Alexandre C.M. Correia,$^{4,5}$
Lalitha Sairam,$^{6}$
Alix V. Freckelton,$^{1}$
Aleyna Adamson,$^{1}$
\newauthor
Isabelle Boisse,$^{2}$
Gavin A.L. Coleman,$^{7}$
Georgina Dransfield,$^{8,9}$
Jo\~ao Faria,$^{10}$
Salomé Grouffal,$^{2}$
\newauthor
Nathan Hara,$^{2}$
Guillaume Hébrard,$^{3}$
Vedad Kunovac,$^{11,12}$
David V. Martin,$^{13}$
Pierre F.L. Maxted,$^{14}$
\newauthor
Richard P. Nelson,$^{7}$
Madison G. Scott,$^{1}$
Owen J. Scutt,$^{1}$
Matthew R. Standing,$^{15}$
\\
$^{1}$School of Physics and Astronomy, University of Birmingham, Edgbaston, Birmingham B15 2TT, UK\\
$^{2}$Aix Marseille Univ, CNRS, CNES, LAM, Marseille, France\\
$^{3}$Institut d’astrophysique de Paris, UMR 7095 CNRS universit\'e Pierre et Marie Curie, 98 bis, boulevard Arago, 75014 Paris, France\\
$^{4}$CFisUC, Departamento de Fisica, Universidade de Coimbra, 3004-516 Coimbra, Portugal\\
$^{5}$IMCCE, UMR8028 CNRS, Observatoire de Paris, PSL Universit\'e, 77avenue Denfert-Rochereau, 75014 , Paris, France\\
$^{6}$Institute of Astronomy, University of Cambridge, Madingley road, Cambridge, CB3 0HA, UK\\
$^{7}$Astronomy Unit, Department of Physics and Astronomy, Queen Mary University of London, Mile End Road, London, E1 4NS, UK\\
$^{8}$Department of Astrophysics, University of Oxford, Denys Wilkinson Building, Keble Road, Oxford OX1 3RH, UK\\
$^{9}$Magdalen College, University of Oxford, Oxford OX1 4AU, UK\\
$^{10}$Observatoire Astronomique de l’Université de Genève, Chemin Pegasi 51, CH-1290 Versoix, Switzerland\\
$^{11}$Department of Physics, University of Warwick, Coventry CV4 7AL, UK\\
$^{12}$Centre for Exoplanets and Habitability, University of Warwick, Gibbet Hill Road, Coventry CV4 7AL, UK\\
$^{13}$Department of Physics and Astronomy, Tufts University, 574 Boston Avenue, Medford, MA 02155, USA\\
$^{14}$Astrophysics Group, Keele University, Keele, Staffordshire, ST5 5BG, UK\\
$^{15}$European Space Agency (ESA), European Space Astronomy Centre (ESAC), Camino Bajo del Castillo s/n, E-28692 Villanueva de la Ca\~nada, Madrid, Spain\\
}

\date{Accepted 2025 July 16. Received 2025 June 17; in original form 2025 April 25}

\pubyear{2025}

\begin{document}
\label{firstpage}
\pagerange{\pageref{firstpage}--\pageref{lastpage}}
\maketitle

\begin{abstract}
Planetary systems orbiting close binaries are valuable testing grounds for planet formation and migration models. More detections with good mass measurements are needed.
We present a new planet discovered during the BEBOP survey for circumbinary exoplanets using radial velocities. We use data taken with the SOPHIE spectrograph at the Observatoire de Haute-Provence, and perform a spectroscopic analysis to obtain high precision radial velocities. This planet is the first radial velocity detection of a previously unknown circumbinary system. The planet has a mass of $0.56$ ${\rm M_{Jup}}$ and orbits its host binary in 550 days with an eccentricity of 0.25. Compared to most of the previously known circumbinary planets, BEBOP-3b has a long period (relative to the binary) and a high eccentricity. There also is a candidate outer planet with a $\sim1400$ day orbital period. We test the stability of potential further candidate signals inside the orbit of BEBOP-3b, and demonstrate that there are stable orbital solutions for planets near the instability region which is where the Kepler circumbinary planets are located.
We also use our data to obtain independent dynamical masses for the two stellar components of the eclipsing binary using High Resolution Cross-Correlation Spectroscopy (HRCCS), and compare those results to a more traditional approach, finding them compatible with one another.
\end{abstract}




\section{Introduction}

Circumbinary exoplanetary systems are a useful test of exoplanet formation and migration models, not only for binary stars but also applicable to single stars. Far from the binary the dynamical influence of the inner binary is negligible and planet formation is expected to be comparable to around single stars. In the inner region of the circumbinary disc, the influence of the binary causes drastically different conditions for nearby bodies compared to single stars; dynamical perturbations induced by the binary cause collisions to be destructive, impairing the growth of planetesimals into protoplanets \citep{meschiari_planet_2012}. This has been shown to mean that circumbinary planet formation is inhibited in a region near the binary \citep{moriwaki_planetesimal_2004,paardekooper_how_2012,pierens_vertical_2021}, however this region is precisely where almost all of the discovered circumbinary planets have been found to exist. These circumbinary exoplanets are therefore likely to have migrated to their current locations, meaning they are an excellent probe of migration theories in general \citep{pierens_formation_2008,coleman_global_2023,coleman_constraining_2024}.

There are currently 16 confirmed circumbinary exoplanets orbiting main-sequence binaries. Of these, 14 were discovered using the transit method \citep[e.g.][]{doyle_kepler-16_2011}, 1 using eclipse timing variations \citep{goldberg_5mjup_2023}, and 1 using the radial velocity method \citep{standing_radial-velocity_2023}. Of these, few are significantly eccentric \citep[the most eccentric is Kepler-34 b with \(e\approx0.18\);][]{welsh_transiting_2012}\footnote{The HD202206 system was discovered through radial-velocity \citep{correia_coralie_2005,benedict_hd_2017} but it is uncertain between a circumbinary brown dwarf and a planet orbiting a Star-Brown dwarf binary, the eccentricity of the "planet" is \(\approx0.22\).} and there are two multi-planet systems: TOI-1338/BEBOP-1 \citep{kostov_toi-1338_2020,standing_radial-velocity_2023} and Kepler-47 \citep{orosz_kepler-47_2012,orosz_discovery_2019}.

Of the transiting circumbinary systems, all but one \citep[Kepler-1647 b,][]{kostov_kepler-1647b_2016} have their innermost planet orbiting very close to the boundary inside which orbits are unstable due to the dynamical influence of the binary \citep{holman_long-term_1999}. This so called "pile-up" is thought to not be solely due to the observational bias of the transit method towards short-period planets \citep{martin_planets_2014,li_uncovering_2016}. The pile-up has been interpreted as a result of migration through a circumbinary disc, and parking around the current location near where the disc gets truncated \citep{penzlin_parking_2021}, and in fact was predicted before being observed \citep{pierens_formation_2008}.

The BEBOP (Binaries Escorted By Orbiting Planets) survey is a radial velocity survey aiming to detect circumbinary planets \citep{martin_bebop_2019}. Its first phase had a focus on single-lined binaries - systems in which there are no obvious spectral lines observed from the secondary star. For a main-sequence binary this typically requires a mass ratio \(q\lesssim 0.3\) \citep{triaud_eblm_2017}. So far BEBOP has made three circumbinary planet detections, with BEBOP-1c \citep{standing_radial-velocity_2023} being the first new planet. The other two detections were confirmations of Kepler-16b \citep{triaud_bebop_2022} and of TIC 172900988b \citep{sairam_new_2024}. This paper, the seventh in the BEBOP series, presents our second circumbinary planet discovery using the radial velocity method, and the first such planet with a significant eccentricity measurement, BEBOP-3b. This newly detected circumbinary planet orbits the eclipsing binary star BD+79\,230 (TIC 289949453).

There is a disadvantage to observing single-lined binaries instead of double-lined binaries. In double-lined binaries, the two sets of radial velocities mean that a model-independent dynamical mass can be derived. In single-lined systems, however, the stellar (and thus the planetary) physical parameters are ultimately dependent on the modeled parameters for the primary star. Recently, \citet{sebastian_eblm_2024} demonstrated the use of High Resolution Cross-Correlation Spectroscopy (HRCCS) on single-lined binaries. This allows the signal of the secondary within the spectra to be retrieved, effectively converting the single-lined binary to a double-lined binary for the purposes of absolute mass determination. \citet{sebastian_eblm_2024} apply the method to the TOI-1338/BEBOP-1 system using spectra from the HARPS and ESPRESSO spectrographs. They obtain a model-independent and more precise mass measurement for both stars in the binary. We repeat the same analysis to obtain dynamical masses for the two components of the BEBOP-3(AB) binary.

In Section \ref{sec:data} we describe the data collection and reduction pipeline. The various analyses performed are described in Section \ref{sec:methods}. We present the results of the radial velocity analysis and a stellar activity analysis in Section \ref{sec:res}. We discuss the results and investigate the orbital stability of the system in Section \ref{sec:diss}, and conclude in Section \ref{sec:conc}.

\section{Radial Velocity Data}\label{sec:data}

The sample selection for the southern component of the BEBOP survey is reported in \citet{martin_bebop_2019}, but the northern sample has not been described much yet. The southern sample was built from a WASP \citep[Wide Angle Search for Planets;][]{pollacco_wasp_2006} false positives flag known as EBLM \citep[Eclipsing Binary -- Low Mass;][]{triaud_eblm_2013,maxted_eblm_2023}. Typically these are Sun-like stars with a stellar companion $< 0.3~\rm M_\odot$, producing transit-shaped primary eclipses. Whilst EBLM-flagged false positives exist for the Northern component of WASP, the record is not as thorough as it was for the south \citep{triaud_constraints_2011}. In the south, the survey benefited from a dedicated spectroscopic follow-up, but in the north, WASP depended on facilities with competitive access and not all false positives were followed-up enough to establish what type they were. This is why the northern BEBOP sample was selected not just from WASP, but also from the KELT survey's \citep[Kilo-degree Extremely Little Telescope;][]{pepper_kilodegree_2007} false positives \citep{collins_kelt_2018}. Systems selected are above the celestial equator, have orbital periods $P_{\rm bin} > 5\,\rm days$, and primary spectral types $>$ F4. The restriction to relatively long period eclipsing binaries was driven by obtainable precision; tight binaries are tidally locked and stars in < 5 day binaries rotate quickly resulting in the spectral lines being so broadened that finding CBPs is very difficult. Coincidentally, the known transiting CBPs are all around similarly "long-period" binaries, potentially as a quirk of their formation process \citep{munoz_survival_2015,martin_no_2015}. Only six systems were common between WASP-North and KELT. Following a short reconnaissance programme in 2018\footnote{Prog.ID 18B.DISC.TRIA}, we created a top priority sub-sample that became the BEBOP-North sample, where SOPHIE spectra typically produced radial velocity uncertainties $< 10 \,\rm m\,s^{-1}$ in 1800s or under.  The BEBOP-3 system was identified from the KELT false positive catalog \citep{collins_kelt_2018}.

Overall, 141 high resolution spectra were obtained between 2018-11-13 and 2024-11-18 with the high resolution \'echelle spectrograph SOPHIE, mounted on the T193 at the Observatoire de Haute-Provence \citep{perruchot_sophie_2008}. Data were acquired under programme ID 18B.DISC.TRIA and 19A.PNP.SANT.

The data were acquired in ObjAB mode where fibre A is placed on the target and fibre B is placed on the sky to enable the removal of solar spectral lines reflected by the Moon. The observations were taken in high resolution (HR) mode with resolution \(R=75\,000\)\footnote{At 5550 \r{A}}. Standard calibrations were taken at the start of the night and at approximately 2~hr intervals throughout to track the instrumental zero-point error.

The SOPHIE radial velocities (RVs) are computed by the SOPHIE Data Reduction System \citep[DRS,][]{bouchy_sophie_2009}, which cross-correlates the spectra with a weighted numerical mask corresponding to a G2-type star \footnote{The BEBOP-3 primary star has a F9 spectral classification, G2 is the most suitable mask within the SOPHIE pipeline.} and subsequently fits a Gaussian to the cross-correlation function (CCF) \citep{pepe_coralie_2002,baranne_elodie_1996}. To enhance the precision of the SOPHIE RV measurements, we applied the optimized procedures described in \cite{heidari_overcoming_2022} and \cite{heidari_sophie_2024}, including: (1) corrections for charge transfer inefficiency in the CCD \citep{bouchy_charge_2009}; (2) correction of template (colour) effects \citep{heidari_sophie_2024}; (3) removal of moonlight contamination by utilising the simultaneous sky spectrum recorded via the second SOPHIE fiber \citep{pollacco_wasp-3b_2008}; (4) correction for long-term instrumental drifts using the RV master constant time series, constructed from so-called ‘constant’ stars monitored each night by SOPHIE \citep{courcol_sophie_2015,heidari_overcoming_2022,heidari_sophie_2024}; and (5) correction for short-term instrumental drifts by interpolating drift measurements at the exact time of each observation, based on frequently measured drifts throughout the night. Radial velocity data are collated into Table \ref{tab:rv_data}.

The 141 spectra obtained have a median SNR\footnote{The signal-to-noise ratio quoted here is that achieved at the center of order 26 (out of orders 0-38), corresponding to 5550 \r{A} (near the center of the V band)} of 72.9 and median radial velocity precision of 2.3 \({\rm m\,s^{-1}}\)
Radial velocity outliers exist. These are dealt with in various ways for the different analyses. Outliers identified are flagged in Table \ref{tab:rv_data}.

\section{Analysis and methods}\label{sec:methods}

\subsection{Spectral analysis}
In \citet{freckelton_bebop_2024}, the 91 SOPHIE spectra available at the time were coadded and the combined spectrum analysed  using the \texttt{PAWS} pipeline. This analysis achieved a signal-to-noise ratio of 555. Based on this signal-noise-ratio being already high enough that model uncertainty is dominant, we do not repeat the analysis from \citet{freckelton_bebop_2024} with extra spectra. Atmospheric parameters for the primary star were determined by applying the \texttt{PAWS} pipeline to this coadded spectrum. The pipeline implements the curve-of-growth equivalent widths (EW) and spectral synthesis methods via \texttt{iSpec} \citep{blanco-cuaresma_modern_2019}. The advantage of combining the two methods sequentially is that the parameters \(T_{\rm eff}\) (Effective temperature), \({\rm log}g\) (Surface gravity), \([\rm Fe/H]\) (Iron abundance), and \(v\sin i\) (projected rotational velocity), as presented in Table \ref{tab:comparison}, are all allowed to freely vary to fit the observations in at least one stage of the pipeline. Using the \texttt{WIDTH} radiative transfer code \citep{sbordone_atlas_2004}, the EW method was used to determine an initial set of atmospheric parameters for the primary star. \texttt{PAWS} then applies the spectral synthesis method to allow a determination of \(v\sin i\), using the initial atmospheric parameters as the start point for the first iteration. Synthetic spectra were generated using these parameters by the use of the \texttt{SPECTRUM} \citep{gray_calibration_1994} radiative transfer code, with the parameters iterated until the synthetic spectrum sufficiently reproduced the observed one, which returned the final atmospheric parameters shown in Table \ref{tab:comparison}. Homogeneity between the two methods was ensured by employing the \texttt{SPECTRUM} \citep{gray_calibration_1994} line list and \texttt{ATLAS} \citep{kurucz_atlas12_2005} model atmosphere set throughout the pipeline.

\subsection{HRCCS measurement of the secondary component}\label{sec:k2}

By combining the large number of SOPHIE observations, we can detect the extremely small contribution of the secondary star ($\sim 0.1\,\%$ for the spectral range of SOPHIE). This in effect turns the single-lined binary into a double-lined binary, which allows us to derive model-independent dynamical masses for both stars. Here we apply the High-Resolution Cross-Correlation Spectroscopy method \citep[HRCCS;][]{snellen_orbital_2010}, which is commonly used to detect exoplanet atmospheres. It has been adapted and fully validated to high-contrast binaries \citep[e.g. TOI-1338/BEBOP-1 and Kepler-16~AB;][]{sebastian_eblm_2024,sebastian_eblm_2025}). The method includes a two step process. First the absorption lines of the primary star are effectively removed from the spectra, second the residual spectra are cross-correlated and combined in the rest-frame of the secondary star. This enhances the SNR of its absorption lines and allows its detection. We follow closely the recipe outlined in \cite{sebastian_eblm_2024}.

We exclude 8 of the 141 SOPHIE spectra from our analysis. Three of which correspond to a wrong pointing and five with very low SNR (<23.5 at 500\,nm), which is 3 times lower than the average SNR (71) for the primary. SOPHIE spectra excluded from the HRCCS analysis are indicated in Table~\ref{tab:rv_data}. The 133 remaining spectra have an average SNR of 73.9 per resolution element at 500\,nm. This means for the faint secondary, the SNR is as low as 0.08 only.

First we split the one dimensional SOPHIE spectra ({\tt s1d}) each into 70 chunks of 4387 pixels. We then shift the spectra into the rest frame of the primary, and perform continuum normalisation, sigma clipping, and singular-value decomposition (SVD) detrending \citep{kalman_singularly_1996}. The latter will effectively remove the spectral lines of the primary star from the series of spectra. The number of excluded SVD components is determined automatically using the effective matrix rank \citep{roy_effective_2007} for each chunk. We set the maximum number of components to 26 and find that the median is 5. This results in a reduced rank of 0.04 (the number of removed components divided by the number of spectra). According to Table~4 in \cite{sebastian_eblm_2024}, we expect that this detrending has removed $\le 3\,\%$ of the secondary's spectral features. The secondary's signal in the data is therefore kept mainly intact.

We exclude a further 11 of the 133 spectra at orbital phases, where the velocity of the secondary's signal differs less than 10\,km\,s$^{-1}$ from the primary's absorption lines. In this way we make sure that possible residuals of the detrending do not interfere with our measurements. We also discharge the 20 bluest chunks (keeping data from 477\,nm and redder) as those add significant noise to the secondary signal. We then use the k-focusing method \citep{sebastian_saltire_2024}, which enhances the cross-correlation function (CCF) of the secondary star in the secondary's rest frame by scanning different semi-amplitudes ($K_2$). This method produces a $K_2$ -- $V_{\rm rest}$ plane by combining all CCFs of the detrended data with an M2-dwarf line mask\footnote{We use the line mask of the HARPS pipeline, which has been publicly released on \hyperlink{https://www.eso.org/sci/software/pipelines/}{https://www.eso.org}}. We sample the map for a velocity range between $-5\,{\rm km\,s^{-1}}$ and $65\,{\rm km\,s^{-1}}$. This range includes the systemic velocity of the binary, measured from the primary star. We choose a step width of $1.5\,\rm km\,s^{-1}$, which takes into account the pixel resolution of the spectrograph. The unknown semi-amplitude of the secondary ($K_2$) is sampled from $55 - 105\,\rm km\,s^{-1}$ in steps of $1.5\,\rm km\,s^{-1}$, making sure the expected semi-amplitude is well covered.

\begin{figure}
    \centering
    \includegraphics[width=\columnwidth]{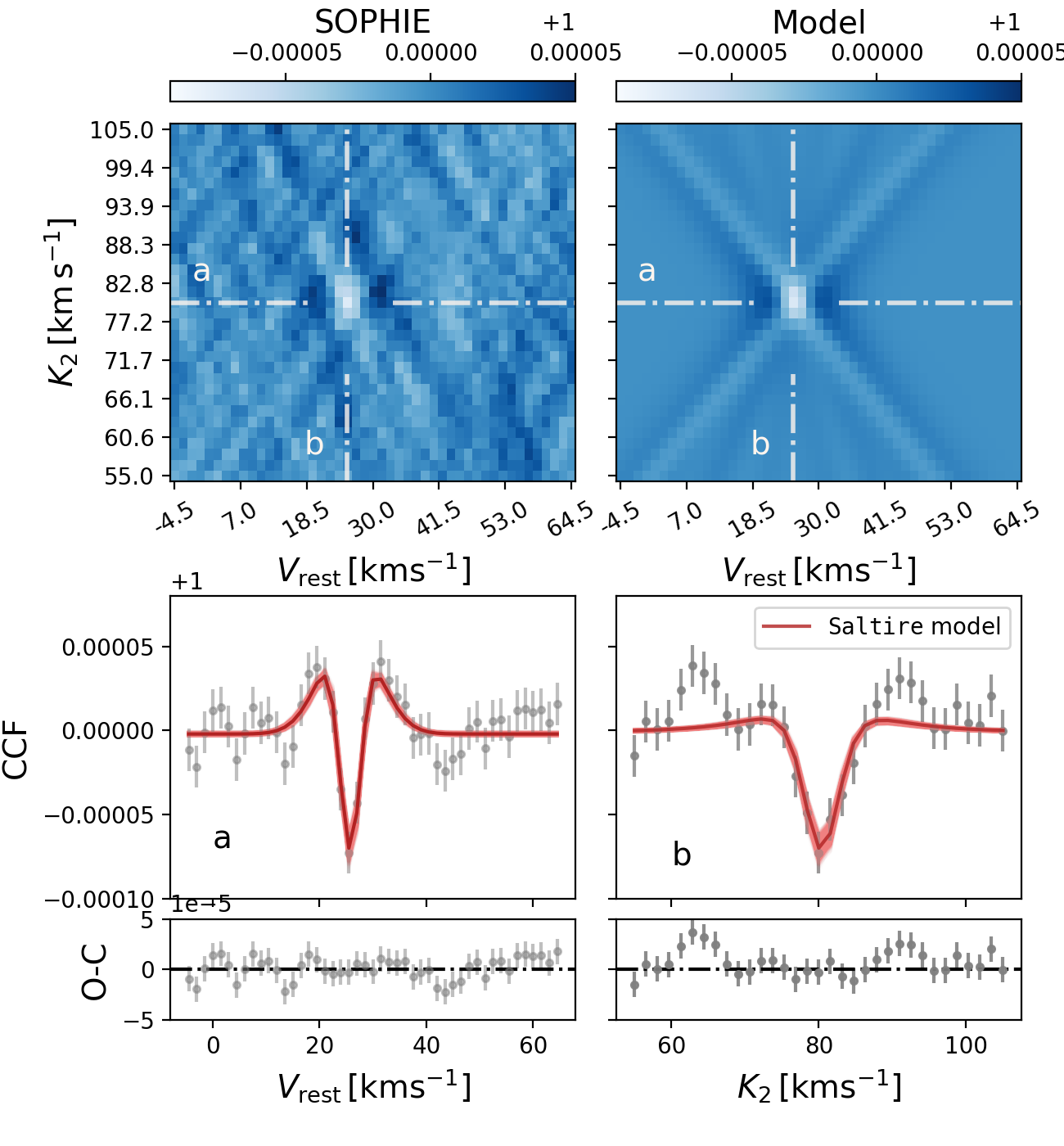}
    \caption{HRCCS 5.5-$\sigma$ detection of the secondary. Upper panel: Cross-correlation map and {\tt Saltire} model for SOPHIE data. White dotted lines show the positions of slices (a,b). Lower panel: Slices a \& b through CCF map at maximum significance. Red line: best-fitting {\tt Saltire} model, red shaded area: $1\sigma$ Uncertainties from the MCMC. Errorbars represent the MCMC jitter term ($\sigma_{\rm jit}$) from the two dimensional fit.}
    \label{fig:HRCCS}
\end{figure}

The resulting CCF map in the $K_2$ -- $V_{\rm rest}$ - plane is shown in Figure~\ref{fig:HRCCS}. The CCF signal of the secondary is detected with a 5.5-$\sigma$ significance at the expected semi-amplitude and systemic velocity. We measure this signal using the {\tt Saltire} model \citep{sebastian_saltire_2024}, which accurately reproduces the CCF signal by fitting a k-focused double-Gaussian function to the data. We estimate the best fitting parameters for $K_2$ and $V_{\rm rest}$ by sampling the data using Markov Chain Monte Carlo (MCMC) with a posterior distribution of 21,000 samples. In Table~\ref{tab:bin_pars}, we present the best parameters, given as the 50th percentile of the MCMC sampling. We then follow \citet{sebastian_eblm_2025} to estimate the systematic uncertainties and randomly draw 50\,\% of the data and measure the best fitting parameters. This is repeated 200 times and the systematic uncertainties are derived as the standard error of the 200 measurements. Results are presented in Section \ref{sec:bin_res} and Tables \ref{tab:comparison} and \ref{tab:bin_pars}.

\begin{table*}
    \centering
    \caption{Stellar and binary parameters from three different analysis methods.}
    \begin{tabular}{l|c|c|c}
        Parameter & HRCCS & RV+Photometry & Spectral \\
        \hline
        \(i_{\rm bin}{\rm (^{\circ})}\) & & \(88.7488\pm0.0092\) & \\
        \(M_{\rm pri}\,({\rm M_{\odot}})\) & \(1.083\pm0.026\) & \(1.163\pm0.043\) & 1.131$\pm$0.051$^{*}$\\
        \(M_{\rm sec}\,({\rm M_{\odot}})\) & \(0.2615\pm0.0039\) & \(0.2731^{+0.0065}_{-0.0064}\) & \\
        \(R_{\rm pri}\,({\rm R_{\odot}})\) & & \(1.414\pm0.017^{**}\) & 1.4107$\pm$0.0089$^{*}$\\
        \(R_{\rm sec}\,({\rm R_{\odot}})\) & & \(0.2811\pm0.0033\) & \\
        \(T_{\rm eff, pri}\,({\rm K})\) & & & \(6033 \pm 100 \)\\
        \(\log g_{\rm pri}\,({\rm dex})\) & & \(4.203\pm0.017\) & \(4.27 \pm 0.09 \) \\
        \({\rm [Fe/H]}\,({\rm dex})\) & & & \(-0.02 \pm 0.10 \)\\
        \(v\sin{i_{\rm pri}}\,({\rm km\,s^{-1}})\) & & & \(2.00 \pm 0.66 \) \\
        \hline
        \multicolumn{4}{l}{$^{*}$ from ISOCHRONE fitting}\\
        \multicolumn{4}{l}{$^{**}$ from \textit{GAIA}}
    \end{tabular}
    \label{tab:comparison}
\end{table*}

\subsection{Joint radial velocity and photometric analysis of the binary}\label{sec:alles}

The goal of the previous section is to obtain precise absolute, dynamical masses. Here we independently analyse the data, using the more standard method when dealing with single-lined eclipsing binaries \citep[e.g.][]{triaud_eblm_2013,davis_eblm_2024}. Our goal in this section is firstly to obtain radii for the two stars and an inclination measurement for the binary, and secondly to be in a position to compare derived (model-dependent) stellar masses, to dynamical (model-independent) stellar masses.

Following \citet{davis_eblm_2024}, we perform a simultaneous joint analysis of both the {\it TESS} \citep[Transiting Exoplanet Survey Satellite;][]{ricker_transiting_2015} photometric data (accessed using the Mikulski Archive for Space Telescopes (MAST) web service\footnote{\url{https://mast.stsci.edu}}
and downloaded using the \texttt{lightkurve} software package \citep{lightkurve_collaboration_lightkurve_2018}) and the SOPHIE spectroscopic data using the {\tt allesfitter} package \citep{gunther_allesfitter_2019, gunther_allesfitter_2021}. {\tt Allesfitter} uses the {\tt dynesty} nested sampling package \citep{speagle_dynesty_2020} and {\tt ellc} \citep{maxted_ellc_2016} for binary system models. All currently available sectors (up to and including Sector 74) of {\it TESS} data are included from the {\it TESS} Science Processing Operations Center (SPOC) pipeline (with 120s cadence) \citep{jenkins_tess_2016}. 

The methods outlined in the remainder of this section are based on those presented in Davis et al. (in prep). The photometric data are chunked into individual datasets to be given to {\tt allesfitter} as this allows for individual hybrid spline baselines to account for any out of eclipse variability. This is more computationally efficient as well as removing unnecessary complexity that comes from employing the use of a gaussian process. Upon visual inspection any eclipse events that were partial, showed larger fluctuations, or showed strong trends in the residuals were removed \footnote{Partial primary eclipse events at BJD 2459022.08, 2459035.13 and 2459405.39, and partial secondary eclipse events at BJD 2459412.10, 2459755.74 and 2459768.92 are removed. Primary eclipse at BJD 2460317.43 and secondary eclipses at BJD 2459015.57 and 2459028.78 removed due to additional photometric variability.} (and the remaining data re-analysed). In this analysis we use 15 of the 19 primary eclipse events. The removed events include three partial eclipses and one eclipse which showed clear additional correlated photometric variability compared to the other primary eclipses. Similarly, we use 14 of the 19 secondary eclipses where of the remove events, three were partial, and two show additional correlated photometric variability which the baseline was unable to account for. We choose to remove the eclipse events with residual correlated variability upon visual inspection to take a cautious approach to the data modeling, mitigating for any systematic bias which could be introduced by including noisier data. All radial velocity outliers identified in Section \ref{sec:analysis} are removed for this analysis.

Before completing the global modelling, the spectroscopic data is first analysed in isolation. This allows for appropriate prior values to be determined for radial velocity parameters related to both the binary and planetary companions. The simultaneous global fit is then completed for the photometric binary companion, and the binary and planetary companions in the spectroscopic timeseries. With the addition of photometric data in the analysis, additional parameters of the system can be determined, including the radius of the stellar companion and the inclination of the binary orbit. Furthermore, by following the parameter derivation method in \citet{davis_eblm_2024}, the mass of the primary star (\(M_{\rm pri}\)) can be determined alongside the mass and radius of the secondary star (\(M_{\rm sec}, R_{\rm sec}\)). These parameters are shown in Table \ref{tab:comparison}. The phase-folded lightcurves are shown in Figure \ref{fig:phase_folded_phot}.

\subsection{Search for planetary signals in the radial velocities}\label{sec:analysis}

The radial velocity timeseries analysis is performed using the "BINARIESmodel" of {\tt kima} \citep{faria_kima_2018,baycroft_improving_2023}. {\tt kima} utilises diffusive nested sampling \citep[DNEST4;][]{brewer_dnest4_2018} to explore the parameter space, and allows for the number of Keplerian signals to be a free parameter. The binary orbit is fit with a Keplerian, using separate priors to the planetary signals. The "BINARIESmodel" is tailored to fitting binary orbits by extending the binary's Keplerian fit with a linear apsidal precession term and by including the radial velocity contribution from general relativity \citep{baycroft_improving_2023}. The parameters used for the Keplerians are \(P,\,K,\,e,\,\omega,\,M_0\), where \(P\) is the orbital period, \(K\) the radial velocity semi-amplitude, \(e\) the orbital eccentricity, and \(\omega\) the argument of periastron. Here \(M_0\) is the mean anomaly at the reference time; for the purpose of reporting the parameters this is converted into a time of pericentre passage \(T_{\rm peri}\). Priors used for the radial velocity analysis with {\tt kima} are shown in Table \ref{tab:priors}.

Outliers are dealt with in a two-stage process. The likelihood calculation uses a Student's t distribution rather than a Gaussian distribution, the wide tails of which allow outliers to avoid skewing the fit while remaining accounted for. The Student's t distribution takes a shape parameter \(\nu\), a small value of which leads to a heavy tail, down-weighting outliers, and a large value results in a near-Gaussian likelihood. Some extreme outliers exist, likely due to incorrect pointings of the telescope. We identify and remove extreme outliers by performing a first analysis with {\tt kima}, identifying any individual datapoints where the ratio of Student's t and Gaussian likelihoods \(L_T/L_N \geq 10^{10}\). These points are removed and the final analysis is then performed, still using a Student's t likelihood to account for any remaining non-extreme outliers. The radial velocity data is made available in Table \ref{tab:rv_data} with the points that we identify as outliers flagged: extreme outliers are flagged with an X; non-extreme outliers are flagged with a Y. Non-extreme outliers are those which have a likelihood ratio \(L_T/L_N \geq 10\) in the final analysis.

\section{Results}\label{sec:res}

\subsection{Absolute physical parameters for the binary}\label{sec:bin_res}

The radial velocity analysis is performed simultaneously for the binary and the potential planet signals searched for. The detection of a planet and criteria related to that are discussed in Section \ref{sec:pl_res}. The orbital parameters of the binary are presented in Table \ref{tab:bin_pars}. We use the values of \(K_2\) from Section \ref{sec:k2} and \(i_{\rm bin}\) from Section \ref{sec:alles} to derive the masses of the two stars.

Table \ref{tab:comparison} shows parameters for the stars obtained from three different analyses, which each give an independent mass measurement. There is a slight tension between the different mass measurements. The measurements from HRCCS (Section \ref{sec:k2}) and joint radial velocity/photometry fit (Section \ref{sec:alles}) have a \(1.7\sigma\) agreement. This difference is not so great, and a thorough study of potential systematic differences between the methods is beyond the scope of this work. We chose to use the mass from HRCCS as it is the least model-dependent, therefore we take \(M_{\rm pri} = 1.083\pm0.026\,M_{\odot}\) and \(M_{\rm sec} = 0.2615\pm0.0039\,M_{\odot}\).

The dynamical mass of each star is therefore obtained to a precision of 2.4\% and 1.5\%. The inclination of \(88.7488\pm0.0092^{\circ}\) is derived from the joint photometric and radial velocity analysis. We report orbital and fit parameters for the stars/binary in Table \ref{tab:bin_pars}, this contains parameters from the radial velocity fit (A few of the parameters depend on results from the HRCCS or photometric fit and these are noted).

\begin{table}
    {\centering
    \caption{Orbital parameters for the BEBOP-3 binary and general fit parameters. Parameters are taken from the {\tt kima} analysis of RVs only unless otherwise specified.The reference time (BJD) used in the analysis is 2459766.975506.}
    \begin{tabular}{l|c|c|c}
        Parameter & Units & Value & Note\\
        \hline
        \(P\) & [days] & \(13.2176657\pm0.0000027\)& \\
        \(K\) & [\({\rm km\,s^{-1}}\)] & \(19.36413^{+0.00091}_{-0.00102}\)& \\
        \(K_2\) & [\({\rm km\,s^{-1}}\)] & \(80.22\pm0.74\) & \({(a)}\)\\
        \(e\) & & \(0.063255\pm0.000054\) & \\
        \(\omega\) & [rad] & \(4.90226^{+0.00097}_{-0.00090}\) & \({(d)}\)\\
        \(\dot{\omega}\) & [\({\rm arcsec\,yr^{-1}}\)] & \(160\pm100\) & \\
        \(\lambda_0\) & [rad] & \(3.66663\pm0.00012\)& \({(d)}\) \\
        \({T_{\rm peri}}\) & [BJD-2450000] & \(9756.1098^{+0.0021}_{-0.0019}\) & \\
        \(M_{\rm pri}\) & [\({\rm M_{\odot}}\)] & \(1.083\pm0.026\) & \({(b)}\) \\
        \(M_{\rm sec}\) & [\({\rm M_{\odot}}\)] & \(0.2615\pm0.0039\) & \({(b)}\) \\
        \(i\) & [deg] & \(88.7488\pm0.0092\)& \({(c)}\) \\
        \(a_{\rm pri}\) & [AU] & \(0.0234853^{+0.0000011}_{-0.0000013}\)& \({(b)}\)\\
        \(a_{\rm sec}\) & [AU] & \(0.09728\pm0.00090\)& \({(b)}\)\\
        \hline
        Jitter & [\({\rm m\,s^{-1}}\)] & \(6.73^{+1.08}_{-0.93}\) & \\
        \(\nu\) & (Student's t shape) & \(7.9^{+29.0}_{-3.6}\) & \\
        \(v_{\rm sys}\) & [\({\rm km\,s^{-1}}\)] & \(24.46146^{+0.00072}_{-0.00075}\)& \\
        \(v_{\rm rest,2}\) & [\({\rm km\,s^{-1}}\)] & \(25.77\pm0.77\)& \({(a)}\) \\
        \hline
    \end{tabular}\\}
    \({(a)}\) Parameter from HRCCS.\\ \({(b)}\) Parameter from combining {\tt kima} results with HRCCS results and using inclination from photometric analysis.\\ \({(c)}\) Parameter taken from joint analysis of RV and photometry.\\ \({(d)}\) The argument of pericentre \(\omega\) and the true longitude at the reference time \(\lambda_0\) are those of the orbit of the primary star around the centre-of mass of the primary-secondary two body orbit, to convert to the parameters for the secondary, \(\pi\) should be subtracted.
    \label{tab:bin_pars}
\end{table}

\subsection{Detection of the planet BEBOP-3b}\label{sec:pl_res}

The trans-dimensional nature of {\tt kima} makes comparing models with different numbers of signals possible from a single analysis. From a single {\tt kima} analysis we can calculate the Bayes Factor for an (n)-planet model compared to a (n-1)-planet model, \(BF_{n,n-1}\). A Bayes Factor over 150 is considered "very strong evidence" for one model over another \citep{trotta_bayes_2008}. We use a maximum of 3 Keplerians, and find that \(BF_{1,0} = 3.4\times10^{13}\), \(BF_{2,1} = 6.0\), and \(BF_{3,2} = 1.1\). This means there is overwhelming evidence for a 1-planet model over a 0-planet model, and weak-to-moderate evidence for a 2-planet model over a 1-planet model. There is no evidence supporting 3 or a higher number of planets, justifying the use of 3 as the maximum, so that computation time is not wasted. For further discussion of the use of Bayesian model comparison with {\tt kima} as a detection metric for planet signals see e.g. \citet{standing_bebop_2022,triaud_bebop_2022,standing_radial-velocity_2023,baycroft_new_2023}. 

\begin{figure}
    \centering
    \includegraphics[width=0.98\columnwidth]{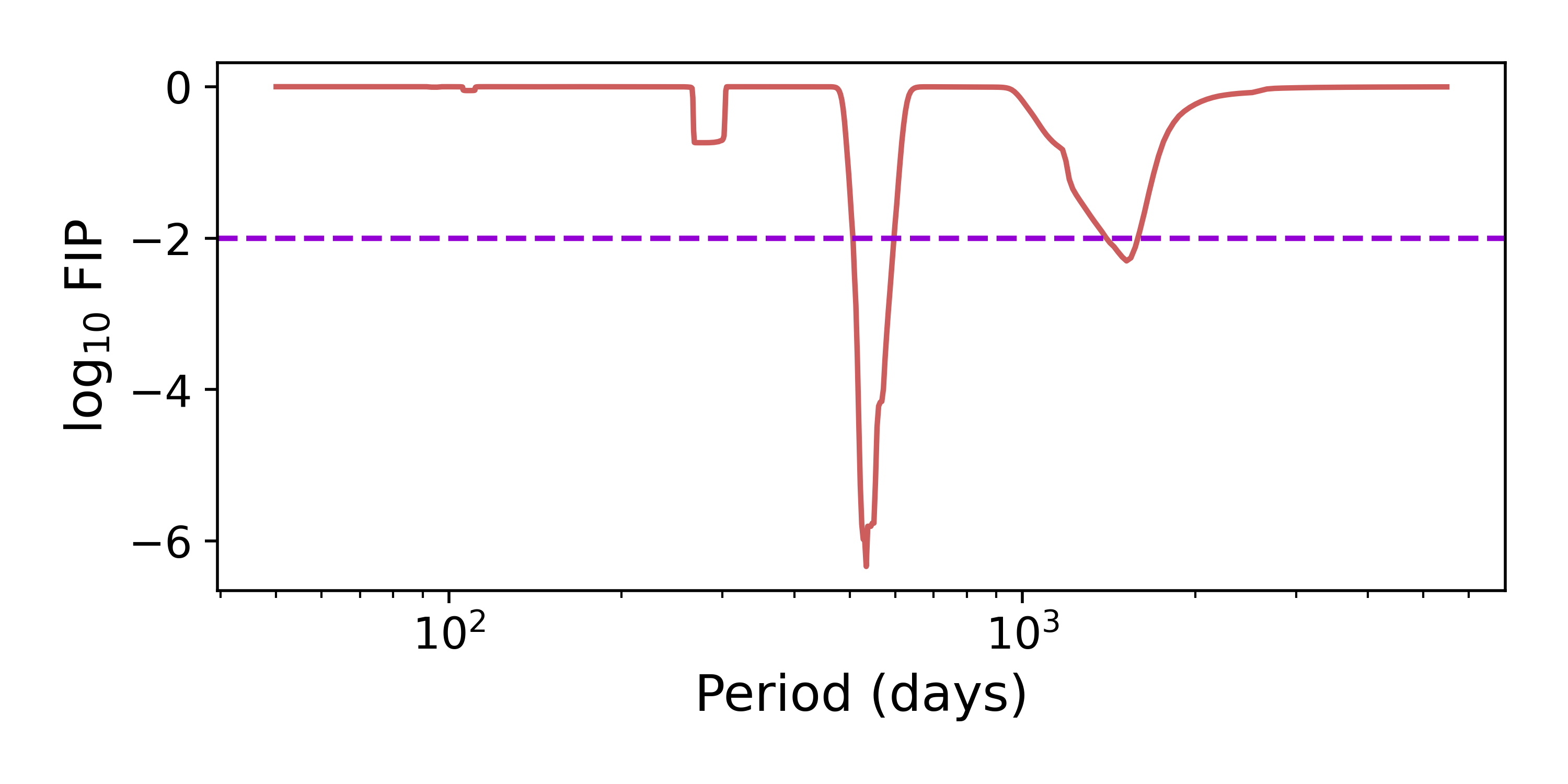}
    \caption{False Inclusion Probability (FIP) periodogram. The dashed purple line is at a \(1\%\) false inclusion probability.}
    \label{fig:FIP}
\end{figure}

While the model comparison favours a 1-Keplerian model, this does not necessarily mean either that the signal is planetary in nature, since some other signal could mimic a Keplerian \citep[e.g.][]{baycroft_new_2023}, or that the parameters of the planet are certain, with aliases and degeneracies being possible. Figure \ref{fig:FIP} shows the False Inclusion Probability (FIP) periodogram \citep{hara_detecting_2022}. The FIP is calculated from the proportion of weighted samples which contain a planet within each period bin. As seen in Figure \ref{fig:FIP} the main detected periodicity is the one between 500-600 days and a second periodicity is seen around 1400 days, just surpassing the \(1\%\) threshold recommended in \citet{hara_detecting_2022}. The FIP has been mathematically shown to be optimal, in the sense that it is the exoplanet detection criterion which maximises true detections for a certain tolerance to false ones~\citep{hara_continuous_2024}. The validation that these signals are not due to stellar activity is presented in Section \ref{sec:activity}.

\begin{figure}
    \centering
    \includegraphics[width=0.9\columnwidth]{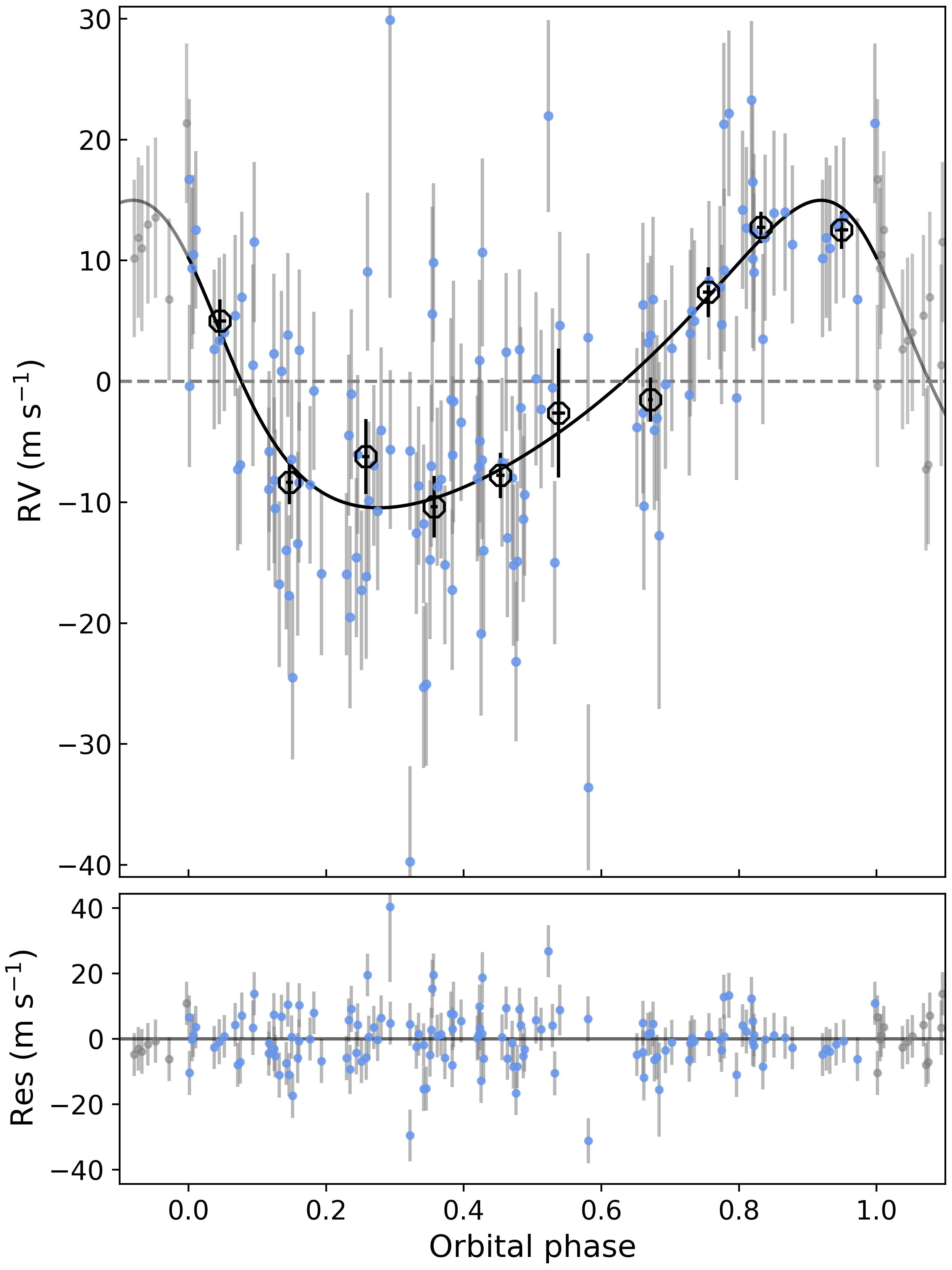}
    \caption{Phased RV plot of BEBOP-3 b, with the residuals below. The parameters for the planet shown are the maximum-likelihood solution. The binned RV data are shown to guide the eye.}
    \label{fig:pl_rv}
\end{figure}

A circumbinary planet with \(P_{\rm pl}=547^{+6.2}_{-7.6}\) days and \(M_{\rm pl}=0.558^{+0.051}_{-0.048}\,M_{\rm Jup}\) is therefore detected, BEBOP-3b. The model favours a moderate eccentricity of \(e=0.247^{+0.077}_{-0.089}\). The maximum-likelihood model for the planetary orbit is shown in Figure \ref{fig:pl_rv} with the individual data shown as well as the residuals below. The binned radial velocities are also shown to guide the eye. The median and \(1\sigma\) uncertainties on the planet parameters are presented in Table \ref{tab:pl_pars}. The distributions and correlations for the posterior samples with the number of planets \(N_p=1\) are shown in Figure \ref{fig:corner_1pl}.

With the use of {\tt kima} we can also report the posterior distributions of the parameters for BEBOP-3b marginalised over the candidate second signal and potential unknown signals. To do this we include the posterior samples from all of the models with different numbers of planets (\(N_p = 1\), \(N_p = 2\), and \(N_p = 3\)). While we prefer to report the parameters from the preferred \(N_p = 1\) model, we also report these marginalised distributions in Table \ref{tab:pl_pars_marg}. These are consistent with the ones reported in Table \ref{tab:pl_pars}, albeit with slightly lower median values of \(K\), \(e\), and \(M\).

\begin{table}
    {\centering
    \caption{Orbital parameters of BEBOP-3 b, note that these are mean orbital elements rather osculating elements. The reference time (BJD) used in the analysis is 2459766.975506.}
    \begin{tabular}{l|c|c|c}
        Parameter & Units & Value & Note\\
        \hline
        \(P\) & [days] & \(547.0^{+6.2}_{-7.6}\) & \\
        \(K\) & [\({\rm m\,s^{-1}}\)] & \(11.8\pm1.1\) & \\
        \(e\) & & \(0.247^{+0.077}_{-0.089}\) & \\
        \(\omega\) & [rad] & \(0.86^{+0.40}_{0.46}\) & \({(b)}\)\\
        \(\lambda_0\) & [rad] & \(2.85\pm0.20\) & \({(b)}\)\\
        \({T_{\rm peri}}\) & [BJD-2450000] & \(9633^{+31}_{-36}\) & \\
        \(M^{(a)}\) & [\({\rm M_{Jup}}\)] & \(0.558^{+0.051}_{-0.048}\) & \({(a)}\)\\
        \(a\) & [AU] & \(1.444\pm0.017\) & \\
        \hline
    \end{tabular}\\}
    \({(a)}\) True mass of the planet under the assumption it is exactly coplanar with the binary.\\ \({(b)}\) the argument of pericentre \(\omega\) and the true longitude at the reference time \(\lambda_0\) are those of the orbit of the binary around the centre-of mass of the planet-binary two body orbit, to convert to the parameters for the planet, \(\pi\) should be subtracted.
    \label{tab:pl_pars}
\end{table}

\begin{figure}
    \centering
    \includegraphics[trim=2cm 1cm 2cm 1cm,width=\columnwidth]{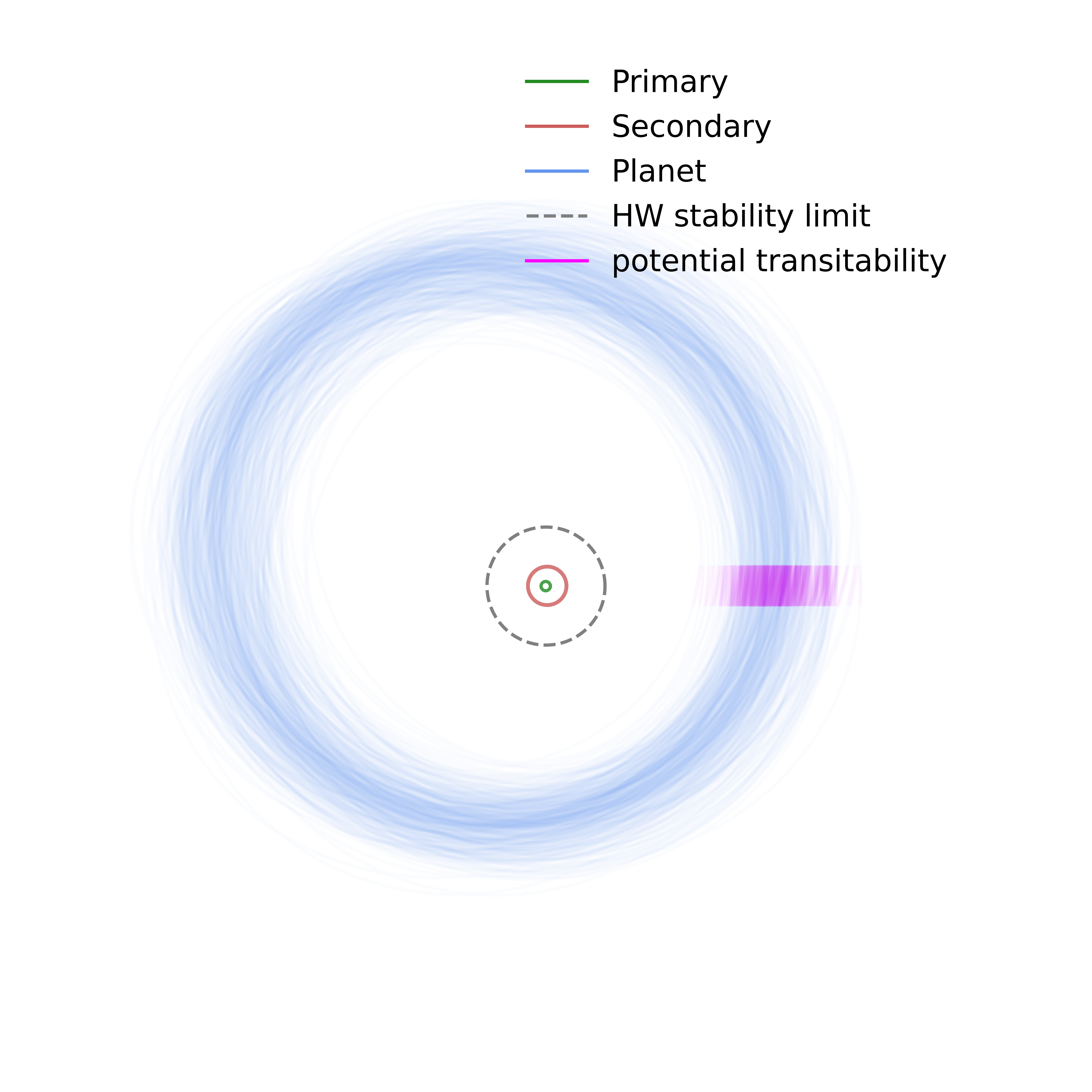}
    \caption{Orbital configuration of BEBOP-3 showing the orbits of each star and the planet for 600 posterior draws. The stability limit is shown as a dashed line, this was calculated given the binary orbital parameters from the maximum-likelihood solution.}
    \label{fig:orbit_plot}
\end{figure}

The orbital configuration of the system is shown in Figure \ref{fig:orbit_plot}. The binary and planet orbits are shown as 600 posterior draws overlain. The region of potential transitability, discussed in Section \ref{sec:trans}, is highlighted in magenta for each planetary orbit shown. The direction to earth is off the right hand side of the diagram. Using the primary mass \(M_{\rm pri} = 1.083\,M_{\rm \odot}\) and the orbital parameters from the maximum likelihood posterior sample, we calculate the stability limit using the equation 3 from \citet{holman_long-term_1999}. This limit is shown on the diagram as a dashed grey line. Defining the scaled semi-major axis as \(a_{\rm sc} = a_{\rm pl}/r_{\rm stab}\), the semi-major axis of the planet's orbit divided by the radius of the stability limit \citep{holman_long-term_1999}, the value for BEBOP-3b is \(a_{\rm sc} = 4.845^{+0.073}_{-0.081}\). Of the known circumbinary planetary systems, all except for Kepler-1647b have their innermost known planet between \(1<a_{\rm sc}<1.8\). This showcases how BEBOP-3b is distinct from the pile-up planets (see Section \ref{sec:eccP_dis}).

\subsection{Further signals}

As mentioned in Section \ref{sec:pl_res}, there is weak-to-moderate evidence for a second planet. The signal corresponds to a planet with \(P\approx1400\) days and \(M\sin{i}\approx0.2\,M_{\rm Jup}\). The posterior distributions and covariances for planet 1 and planet 2 within the \(N_p=2\) posteriors are shown in figures \ref{fig:corner_2pl_1} and \ref{fig:corner_2pl_2}.

\begin{figure}
    \centering
    \includegraphics[width=\columnwidth]{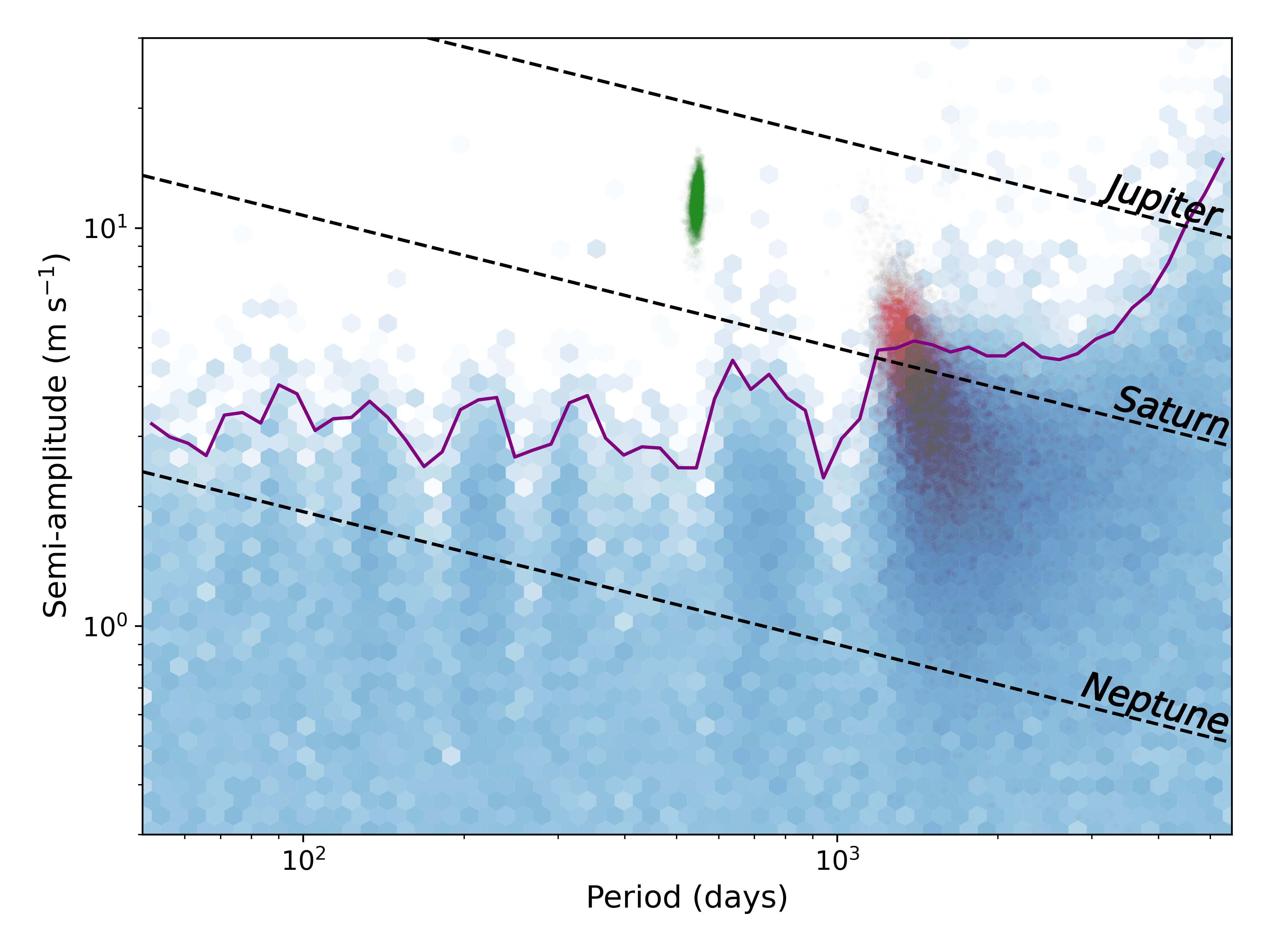}
    \caption{Detection limit for BEBOP-3. In blue, the posterior density of the {\tt kima} run forced to fit a signal, with planet b removed. The posterior distribution from the {\tt kima} radial velocity analysis is shown in green for planet-b (\(N_p = 1\) model)} and that for the candidate outer signal (\(N_p = 2\) model) is shown in red. Dashed lines show expected signals of solar-system mass planets.
    \label{fig:detlim}
\end{figure}

We compute a detection limit to constrain potential other planets in this system. These are computed as described in \citet{standing_bebop_2022,standing_radial-velocity_2023}. The maximum-likelihood signal of the detected planet is removed, and then the residual data is analysed with the number of planets that {\tt kima} fits for being fixed to one. The binary signal is not removed, and is then fit in this second analysis in the same way as before. This analysis then returns a posterior distribution of orbits which are consistent with the data but were not detected. The resulting posterior density is shown in Figure \ref{fig:detlim} along with the 99\% confidence detection limit. Sensitivity is reached down to Neptune-mass planets at short orbital periods and to Saturn-mass planets at long orbital periods.

In the detection limit in Figure \ref{fig:detlim}, the posterior distributions for planet-b and this outer candidate signal are also displayed. The planet-b distribution is that from the 1-planet posteriors from the {\tt kima} analysis in Section\,\ref{sec:pl_res} (shown in Figure\,\ref{fig:corner_1pl}). The candidate's distribution is that from the 2-planet model (shown in Figure\,\ref{fig:corner_2pl_2}.) The distribution for the outer signal is majoritarily below the detection limit because the signal is not large enough (or the data not precise/accurate enough) to be detected/refuted. 

Further observations are ongoing to confirm/refute this candidate signal. In Section \ref{sec:activity} the highest activity signal is found at \(\sim1000\) days. Obtaining more data will also increase the significance and the precision of the activity peaks, and help to ascertain if the candidate signal is from activity or not.

\subsection{Stellar activity analysis}\label{sec:activity}

\begin{figure}
    \centering
    \includegraphics[width=\columnwidth]{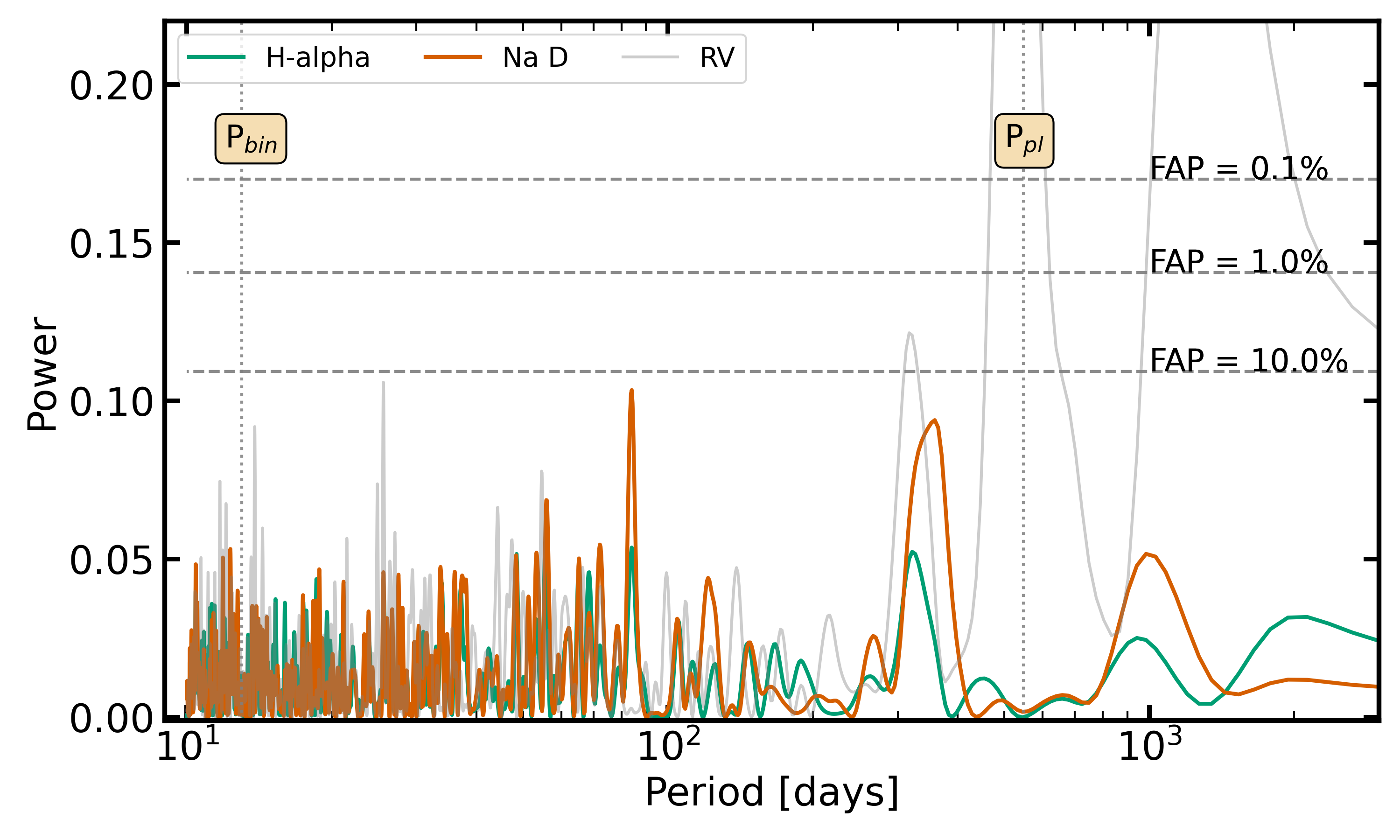}
    \caption{Periodograms of the activity indicators H-alpha and NaD are presented in orange and green, respectively with the periodogram of the radial velocities (with the binary solution removed) plotted in grey. Grey dotted vertical lines denote the binary period and planetary period. Grey horizontal dashed lines represent FAP levels at 10\%, 1\%, and 0.1\%}
    \label{fig:activity_periodogram}
\end{figure}

To investigate the periodic variations in the chromospheric activity indicators, we performed a Generalised Lomb-Scargle periodogram analysis of the H-alpha and Na D indices (similar to \citealt{standing_radial-velocity_2023}). These indices exhibit the highest signal-to-noise ratio in our dataset, making them the most reliable tracers of periodic signals. Other indicators, such as Ca I, He I, and Ca II, displayed significant noise and did not reveal clear periodicities. The analysis was conducted on all the observations spanning a time baseline of 2197 days except those deemed outliers. The same outliers were removed as in the {\tt kima} radial velocity analysis with which the planet was identified, these are marked with an X in Table \ref{tab:rv_data}. The periodograms are shown in Figure \ref{fig:activity_periodogram} along with False Alarm Probability (FAP) levels. The FAP is calculated by determining the power levels corresponding to a given probability threshold, which quantifies the likelihood that a detected signal is due to random noise rather than a true periodic signal. In this analysis, the FAP levels are set at 10\%, 1\%, and 0.1\%, and their corresponding power thresholds are computed from the periodogram to assess the significance of the detected peaks. 

For the H-alpha index, the periodogram identifies a peak at a frequency of 0.0179 $\pm$ 0.0001 d$^{-1}$, corresponding to a period of 55.57 $\pm$ 0.35days.  The maximum power of the detected signal is 0.05. However, when compared with the false alarm probability levels, this power remains below the 10\% significance threshold, indicating that the detected periodicity is not statistically significant. The weighted mean of the H-alpha indices is 0.1868, with an RMS scatter of 0.006.

A similar analysis of the Na D index reveals a dominant period of 84.07 $\pm$ 0.72 days. The power of this signal is $\sim$0.1, which also remains below the 10\% FAP significance threshold, suggesting that the observed periodicity is not robust. The weighted mean of the Na D indices is 0.4365, with an RMS scatter of 0.0195. For comparison, the periodogram of the radial velocities, with the binary signal removed, is shown in the background of Figure \ref{fig:activity_periodogram}.

Although both activity indices exhibit periodic variations at similar timescales, the lack of statistical significance implies that these detections will require further observations and improved data sampling to confirm the nature of the signal. Neither H-alpha periodicity nor Na D periodicity are close to the strong signal seen around 550 days in the radial velocities. As such it is unlikely stellar photospheric processes are causing the observed radial-velocity modulation. We therefore interpret it as planetary.

\section{Discussion}\label{sec:diss}

\subsection{Stability analysis}\label{sec:stability}

The BEBOP-3(AB)b system is hierarchical ($a_{\rm bin} / a_{\rm b}  \ll 1$), this condition typically results in stability as it avoids mean-motion resonance overlap between the planet and binary, many numerical studies find these orbits to be stable \citep[e.g.][]{holman_long-term_1999,doolin_dynamics_2011}. Therefore if there are no planets interior to its orbit the stability of BEBOP-3b should be assured \citep[e.g.][]{correia_secular_2016}. Motivated by the population of known circumbinary planets located near the stability limit, we investigate the stability of the system when including orbits interior to that of planet b. We simulate the BEBOP-3 system using two close-in stars in a nearly circular orbit ($a_{\rm bin} = 0.12$~au, $e_{\rm bin} = 0.06$, Table~\ref{tab:bin_pars}), together with an outer planet further away in an eccentric orbit ($a_{\rm b} = 1.44$~au, $e_{\rm b} = 0.28$\footnote{The maximum-likelihood parameters from the {\tt kima} analysis are used for the stability analysis which while within the uncertainties presented in Tables \ref{tab:bin_pars} and \ref{tab:pl_pars}, do not always correspond exactly to the mean values of the posterior reported.}, Table~\ref{tab:pl_pars}).

In order to get a reliable and comprehensive view of the stability of potential inner planets in the system, we performed a global frequency analysis \citep{laskar_chaotic_1990, laskar_frequency_1993} in the same way as achieved for other circumbinary planetary systems \citep[eg.][]{correia_coralie_2005, standing_radial-velocity_2023}.
The system is integrated on a regular 2D mesh of initial conditions in the vicinity of the best fit (Tabs.~\ref{tab:bin_pars} and~\ref{tab:pl_pars}).
We used the symplectic integrator SABAC4 \citep{laskar_high_2001}, with a step size of $0.001$~yr and general relativity corrections.
Each initial condition is integrated for 5000~yr, and a stability indicator, $\Delta = |1-n_{b'}/n_b|$, is computed. Here, $n_b$ and $n_b'$ are the main frequency of the mean longitude of the planet over 2500~yr and 5000~yr, respectively, calculated via the frequency analysis \citep{laskar_frequency_1993}. The results are reported in color, where yellow represents strongly chaotic trajectories with $\Delta > 10^{-2}$, while extremely stable systems for which $\Delta < 10^{-8}$ are shown in purple/black. 
Orange indicates the transition between the two, with $\Delta \sim 10^{-4}$.

\begin{figure*}
    \centering
    \includegraphics[width=\textwidth]{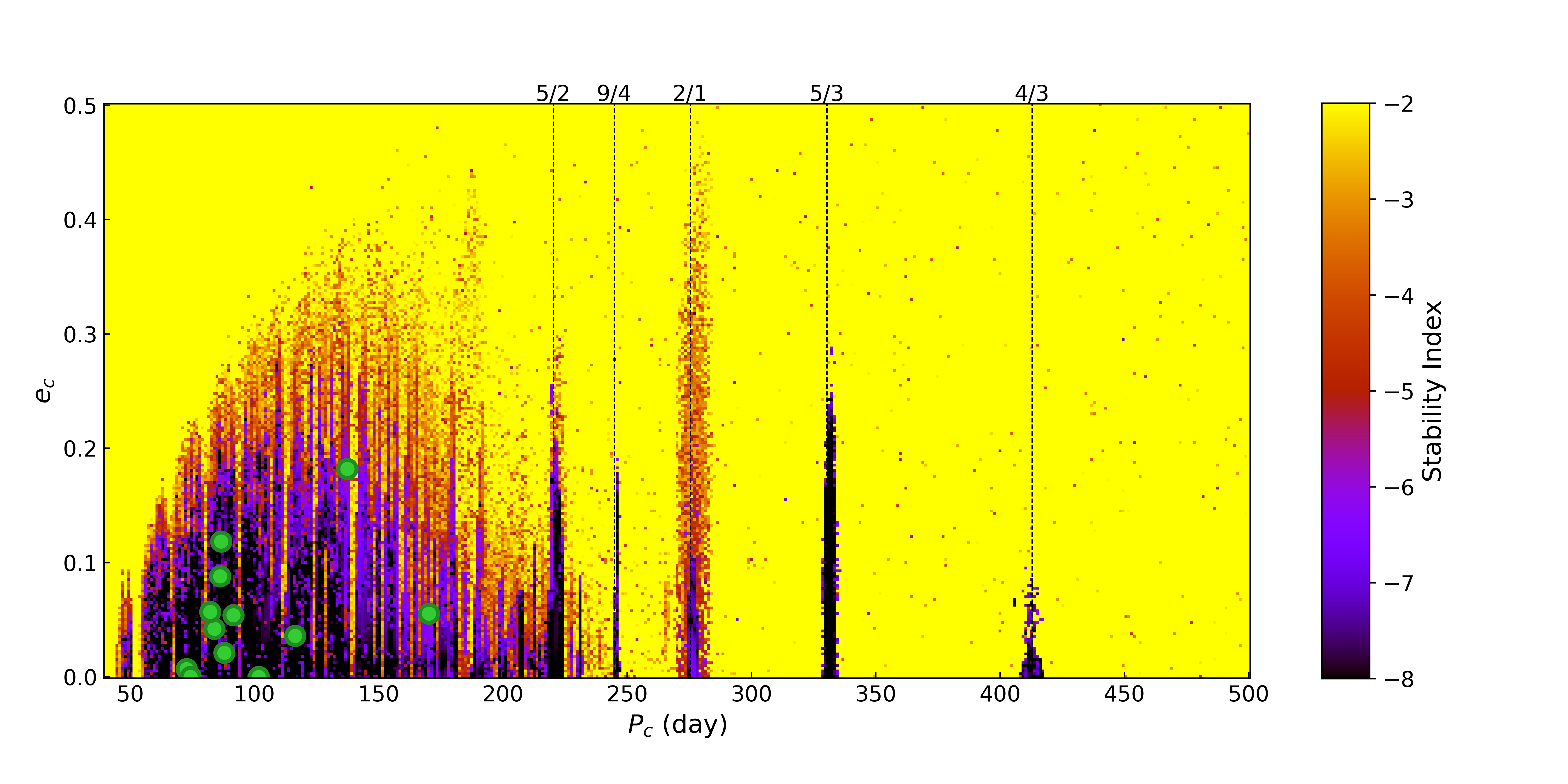}
    \vspace{-3.em}
    \caption{Stability analysis of an inner planet in the BEBOP-3 system assuming $m_\mathrm{inner} \sim 10$~M$_\oplus$ and coplanar orbits. For fixed initial conditions (Tabs.~\ref{tab:bin_pars} and~\ref{tab:pl_pars}), the parameter space of the system is explored by varying the orbital period $P_\mathrm{inner}$ and the eccentricity $e_\mathrm{inner}$ of planet-c. The step size is $1.15$~day in orbital period and $0.0025$ in eccentricity. For each initial condition, the system is integrated over 5000~yr and a stability indicator is calculated which involved a frequency analysis of the mean longitude of the inner planet (see text). Chaotic diffusion is indicated when the main frequency of the mean longitude varies in time. Yellow points correspond to highly unstable orbits, while purple points correspond to orbits which are likely to be stable on a billion-years timescales. The dashed black lines indicate mean-motion resonances with the planet BEBOP-3b. The green circles represent the locations of the innermost circumbinary planet in the other known systems, scaled to the parameters of the BEBOP-3 binary.}
    \label{fig:SF3}
\end{figure*}

We test the possibility of an inner planet, between the binary stars and planet-b, with a radial-velocity signature of 1~\({\rm m\,s^{-1}}\). Such a planet would correspond to a mass $m_\mathrm{inner} \sim 10 $~M$_\oplus$.  
We assume the best fit solution for the already known bodies in the system (Tabs.~\ref{tab:bin_pars} and~\ref{tab:pl_pars}), coplanar orbits, and vary the orbital period and the eccentricity of a putative inner planet (Figure~\ref{fig:SF3}).
We observed that close to the binary ($50 \lesssim P_\mathrm{inner} \lesssim 200$~days), the system is mostly stable with small eccentricity ($e_\mathrm{inner} \lesssim 0.1-0.2$), except close to some mean-motion resonances between the putative inner planet and the binary.
Interestingly, as we move further away ($P_\mathrm{inner} \gtrsim 200$~days), we observed the exact opposite behaviour: the system becomes globally unstable, but stability is still possible, even with moderate eccentricities, near some mean-motion resonances between the inner planet and planet-b.

In Figure\,\ref{fig:SF3}, the location of the known innermost planets of all the known circumbinary systems (excluding Kepler-1647b) are shown. The orbital period has been scaled by the ratio of the orbital period of the respective binaries. All of these points lie in the region of stability, and 6 of these 12 planets have a mass measurement (or non-measurement) consistent with being below the mass of Neptune, which is the approximate sensitivity limit in this region for BEBOP-3. This suggests that an additional planet consistent with the current population would remain undetected in the data collected thus far.

\subsection{An eccentric long-period circumbinary planet}\label{sec:eccP_dis}

Of the circumbinary planetary systems discovered by the transit method or radial velocities, only Kepler-1647 \citep{kostov_kepler-1647b_2016} does not contain a planet orbiting close to the instability limit \citep{martin_planets_2014}. BEBOP-3 joins Kepler-1647 as the second circumbinary planetary system in this category. Additionally, some circumbinary planets have been found further from the stability limit, detected with other methods such as microlensing \citep{bennett_first_2016} and eclipse timing variations \citep[e.g.][]{qian_detection_2011}.

In the cases of Kepler-1647 and BEBOP-3 it is unclear whether there are additional, currently undetected, planets orbiting closer to the binary or whether these circumbinary systems followed a different formation and evolutionary pathway. The stability analysis in Section \ref{sec:stability} shows that there is a significant stable region exactly where it would be expected for a planet comparable to the Kepler planets to exist. In addition, the inner planets of circumbinary systems regularly 
appear to have a low mass (e.g. Kepler-47b \citep{orosz_kepler-47_2012}, TOI-1338/BEBOP-1b \citep{standing_radial-velocity_2023}). In the case of TOI-1338/BEBOP-1b, despite 184 spectra, the planet remains undetected in the RVs alone \citep{standing_radial-velocity_2023}. Based on these considerations it could be expected that either a planet below the radial velocity detection threshold is present there, or that a planet used to be present there but has since become destabilised.

\begin{figure}
    \centering
    \includegraphics[width=\columnwidth]{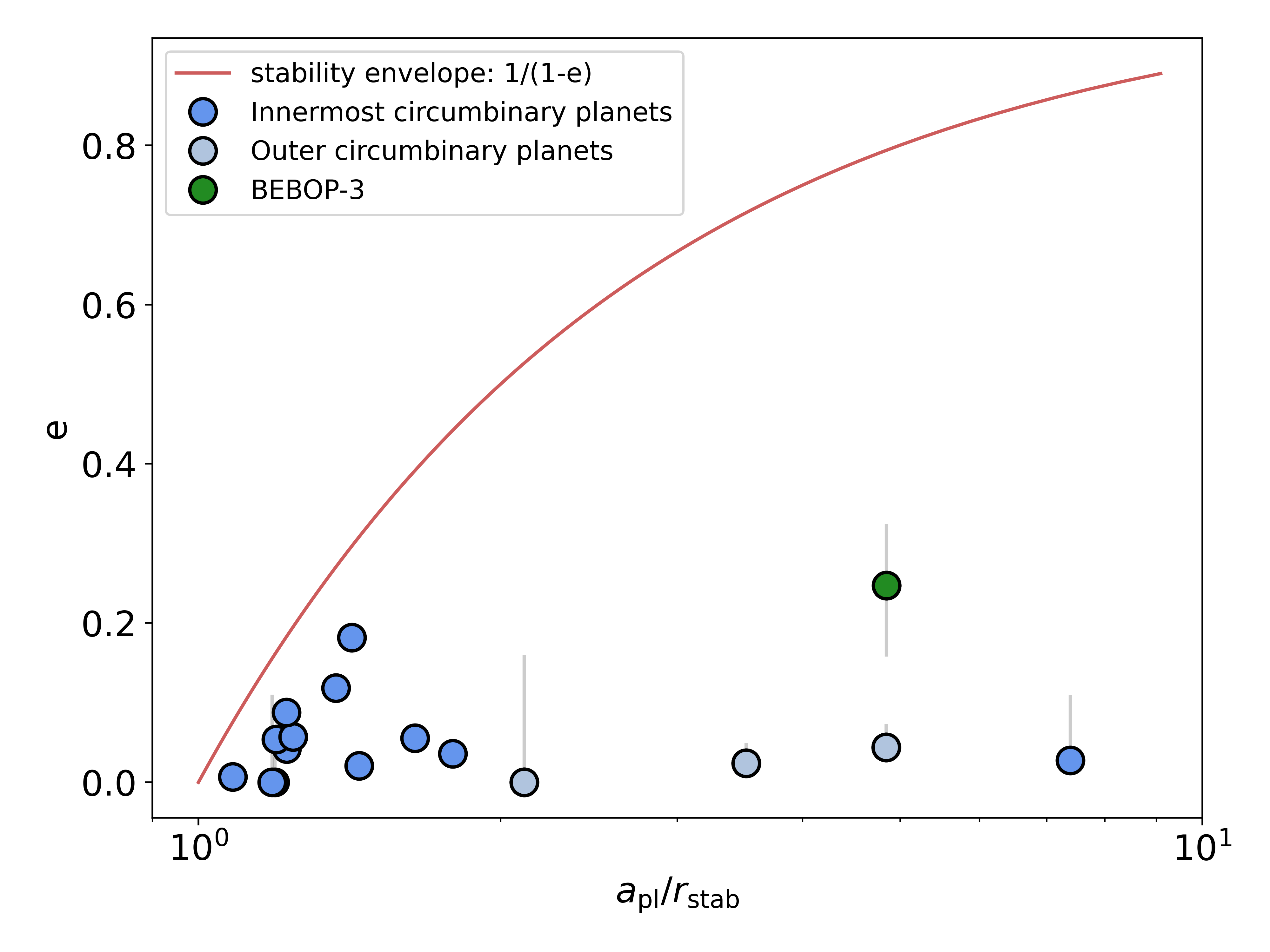}
    \caption{Eccentricity distribution of the population of currently known Circumbinary exoplanets. The x-axis is the semi-major axis of the planets scaled by the Holman-Wiegert stability radius for each binary. The green point represents BEBOP-3 b. The red curve is an indicative stability envelope above which orbits are expected to be unstable.}
    \label{fig:eccpop}
\end{figure}

BEBOP-3b has a marginal detection of eccentricity. It is the only circumbinary planet with a non-zero measured eccentricity from the radial velocity method so far. Figure \ref{fig:eccpop} displays the location of BEBOP-3b compared to the other circumbinary planets in eccentricity and semi-major axis scaled by the radius of the stability limit \citep[calculated from equation 3 in ][]{holman_long-term_1999}. The red curve shows an approximate stability envelope above which a planet's orbit would cross into the instability zone. The known circumbinary planets are mostly clustered in the bottom left corner of this plot, near the stability limit and at low eccentricities. The planets with higher eccentricities have a slightly larger semi-major axis to compensate for this, leading to a pattern of planets tracing just below the stability envelope. As is visible in Figure \ref{fig:eccpop}, BEBOP-3b is distinct from the other known circumbinary planets in being both eccentric and far from crossing into the instability region. 

Despite appearing somewhat special among circumbinary planets, there are a few other considerations/caveats: 
\begin{itemize}
    \item Many gas giants orbiting single stars occupy eccentric orbits \citep[e.g.][]{winn_occurrence_2015}, and compared to them BEBOP-3b is not abnormal.
    \item This is only the second RV-discovered circumbinary planet, and the second longest orbital period; only more detections can tell whether its current parameters are unusual or not.
    \item Eccentricity measurements of planetary signals from radial velocities can sometimes be unreliable. This is made worse when there are undetected planets in the system \citep{hara_bias_2019}, in particular when near orbital period commensurabilities \citep{anglada-escude_how_2010}. Comparing the eccentricity distributions in Figures \ref{fig:corner_1pl} and \ref{fig:corner_2pl_1} show that including the outer planet candidate slightly decreases the significance of the eccentricity measurement. More data is required to assess both the significance of the second planet and of the eccentricity measurement. 
    \item The prior used for the eccentricity is that from \citet{kipping_parametrizing_2013} which is tailored to planets around single-stars and may not be suitable for circumbinary planets. 
\end{itemize}

More circumbinary planet detections and better characterisation of the existing planets are needed to determine whether there is a population of lone long-period circumbinary planets, and also to better constrain the eccentricity distribution. 

\citet{coleman_constraining_2024} investigated the formation history of circumbinary planets in the TOI-1338/BEBOP-1 system, which has similar stellar masses and binary orbital period to BEBOP-3 (see Table \ref{tab:catalogue}. The parameters of BEBOP-3b presented here appear consistent with the population of planets formed in their model.

\subsection{Transitability prospects}\label{sec:trans}

Circumbinary planets, if misaligned with the inner binary, will undergo nodal precession of their orbit. This means that most (\(\approx 95\%\)) of circumbinary planets orbiting eclipsing binaries are expected to eventually come into a transitable configuration \citep{martin_circumbinary_2015}. Transitability means that the planet crosses in front of the path of the binary orbit and might therefore transit in front of one or both of the stars, however in a given conjunction the planet still might happen to not transit. More thorough discussions can be found in  \citet{martin_circumbinary_2015} and \citet{martin_circumbinary_2017}.

From our radial velocity solution for BEBOP-3b, we cannot assess the mutual inclination between the orbits of the planet and binary, or the position of the planet in its nodal precession cycle. We therefore consider times of "possible transitability", when the planet is near inferior conjunction. If the planet is the correct orientation to be able to transit, it would have to do so during these times.

We take 1000 posterior samples for the binary and planet parameters and calculate the times of possible transitability for each sample. We then plot a histogram of these times, giving a posterior probability distribution of times of possible transitability for each conjunction of the planet. These are compared with the times that the system has been observed with {\it TESS} (TIC 289949453) \citep{ricker_transiting_2015}. Based on this, even if the planet did transit at any of these conjunctions, it is highly unlikely to be present in any {\it TESS} data. We do note that the orbital configuration of the planet, as shown in Figure \ref{fig:orbit_plot} is optimal for a planetary transit as the transit would occur very near pericentre where the distance between planet and star is smallest.

\begin{figure*}
    \centering
    \includegraphics[width=2\columnwidth]{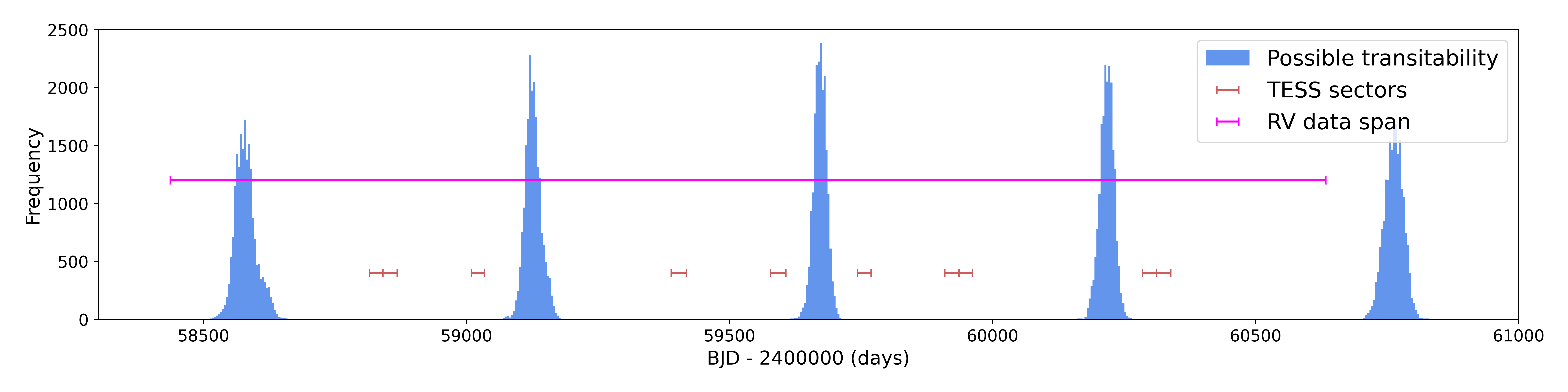}
    \caption{Distributions of potential transitability of BEBOP-3 b from the radial velocity fit. The TESS sectors in which BEBOP-3 was observed are shown, along with the range of time over which the radial velocities were taken.}
    \label{fig:transits}
\end{figure*}

\subsection{The BEBOP catalogue of circumbinary planets with radial velocity measurements}

BEBOP-3b is now the fourth published circumbinary planet detected using radial velocities
. We introduce here the BEBOP catalogue of circumbinary planets with radial velocity measurements and the rationale for the numbering. The catalogue is shown in table \ref{tab:catalogue}. This catalogue is intended to collect all of the circumbinary planetary systems with parameters measured using radial velocities in one place. The catalogue does include circumbinary planets that were detected first in transit and might include planets detected by other teams using radial velocities. For each system we give a BEBOP catalogue entry number without presuming to replace the names.

Kepler-16b and TIC 172900988b were first detected in transit \citep{doyle_kepler-16_2011,kostov_tic_2021} and subsequently detected in radial velocities \citep{triaud_bebop_2022,sairam_new_2024}. Based on the timing of their detections relative to BEBOP-1c \citep{standing_radial-velocity_2023} we number them 0 and 2 in the BEBOP catalogue leaving the new planet presented here as number 3. Note that TOI-1338b which was detected in transit \citep{kostov_toi-1338_2020} has not yet been detected in radial velocities\footnote{There is a photodynamical mass measurement presented in \citet{wang_photo-dynamical_2024} but it is unclear to what extent the measurement is influenced by the radial velocities.} so the inner planet is only given an upper mass-limit in the BEBOP catalogue.

\section{Conclusion}\label{sec:conc}

In this work we present the radial velocity detection of a circumbinary planet BEBOP-3b. This planet holds a distinct position relative to the known population of circumbinary planets orbiting main-sequence binaries. BEBOP-3b orbits with a long period relative to the binary and is on a moderately eccentric orbit. A dynamical analysis reveals that there is ample space for another undetected planet near the stability limit, so BEBOP-3 may not be so outlying relative to the known population. There is an additional longer-period candidate signal with insufficient evidence to claim as a detection. We extracted stellar activity indices, and none show variability on the radial-velocity modulation we observed. We consequently interpret this modulation as being produced by Doppler reflex motion of a circumbinary planet. The absolute dynamical mass of the stars are calculated using high resolution cross-correlation spectroscopy. The planet's minimum mass as derived from the radial velocities is likely close to its true mass, assuming coplanarity. Should the inclination of its orbit become known, BEBOP-3b would also have an absolute mass.

\section*{Acknowledgements}

These scientific results would not have been possible without a generous allocation of telescope time by France Programme National de Planétologie, and by the dedicated staff working at the Observatoire de Haute-Provence, particularly during the COVID pandemic. We also would like to thank the help of other observers part of the timeshare agreement between groups using the SOPHIE instruments. 
This research leading to these results is supported by a grant from the European Research Council (ERC) under the European Union's Horizon 2020 research and innovation programme (grant agreement n$^\circ$ 803193/BEBOP) and by the ERC/UKRI Frontier Research Guarantee programme (EP/Z000327/1/CandY). This work was supported by the "Programme National de Plan\'etologie" (PNP) of CNRS/INSU co-funded by CNES and the "Action Th\'ematique de Physique Stellaire" (ATPS) of CNRS/INSU PN Astro co-funded by CEA and CNES. ACMC acknowledges support from FCT - Funda\c{c}\~ao para a Ci\^encia e a Tecnologia, I.P., Portugal, through the projects 2024.01252.BD, CFisUC (UIDB/04564/2020 and UIDP/04564/2020, with DOI identifiers 10.54499/UIDB/04564/2020 and 10.54499/UIDP/04564/2020, respectively). AVF acknowledges the support of the IOP through the Bell Burnell Graduate Scholarship Fund. RPN acknowledges funding from the STFC grant ST/X000931/1.
Computations for this research were performed using the University of Birmingham's BlueBEAR HPC service \href{http://www.birmingham.ac.uk/bear}{http://www.birmingham.ac.uk/bear}. The stability maps were performed at the Oblivion Supercomputer at the University of \'Evora (\href{https://oblivion.hpc.uevora.pt}{https://oblivion.hpc.uevora.pt}).

\section*{Data Availability}

The {\it TESS} data is available to download via the MAST portal (\href{https://mast.stsci.edu}{https://mast.stsci.edu}). Spectroscopic data are available on the SOPHIE archive (\href{http://atlas.obs-hp.fr/sophie/}{http://atlas.obs-hp.fr/sophie/}), and were obtained under Prog.ID 18B.DISC.TRIA and 19A.PNP.SANT. Reduced radial velocities are found in Table \ref{tab:rv_data}, a full version of which can be found online.



\bibliographystyle{mnras}
\bibliography{BEBOP3} 

\begin{thebibliography}{}
\makeatletter
\relax
\def\mn@urlcharsother{\let\do\@makeother \do\$\do\&\do\#\do\^\do\_\do\%\do\~}
\def\mn@doi{\begingroup\mn@urlcharsother \@ifnextchar [ {\mn@doi@} {\mn@doi@[]}}
\def\mn@doi@[#1]#2{\def\@tempa{#1}\ifx\@tempa\@empty \href {http://dx.doi.org/#2} {doi:#2}\else \href {http://dx.doi.org/#2} {#1}\fi \endgroup}
\def\mn@eprint#1#2{\mn@eprint@#1:#2::\@nil}
\def\mn@eprint@arXiv#1{\href {http://arxiv.org/abs/#1} {{\tt arXiv:#1}}}
\def\mn@eprint@dblp#1{\href {http://dblp.uni-trier.de/rec/bibtex/#1.xml} {dblp:#1}}
\def\mn@eprint@#1:#2:#3:#4\@nil{\def\@tempa {#1}\def\@tempb {#2}\def\@tempc {#3}\ifx \@tempc \@empty \let \@tempc \@tempb \let \@tempb \@tempa \fi \ifx \@tempb \@empty \def\@tempb {arXiv}\fi \@ifundefined {mn@eprint@\@tempb}{\@tempb:\@tempc}{\expandafter \expandafter \csname mn@eprint@\@tempb\endcsname \expandafter{\@tempc}}}

\bibitem[\protect\citeauthoryear{Anglada-Escudé, López-Morales  \& Chambers}{Anglada-Escudé et~al.}{2010}]{anglada-escude_how_2010}
Anglada-Escudé G.,  López-Morales M.,   Chambers J.~E.,  2010, \mn@doi [The Astrophysical Journal] {10.1088/0004-637X/709/1/168}, 709, 168

\bibitem[\protect\citeauthoryear{Baranne et~al.,}{Baranne et~al.}{1996}]{baranne_elodie_1996}
Baranne A.,  et~al., 1996, Astronomy and Astrophysics Supplement, v.119, p.373-390, 119, 373

\bibitem[\protect\citeauthoryear{Baycroft, Triaud, Faria, Correia  \& Standing}{Baycroft et~al.}{2023a}]{baycroft_improving_2023}
Baycroft T.~A.,  Triaud A. H. M.~J.,  Faria J.,  Correia A. C.~M.,   Standing M.~R.,  2023a, \mn@doi [Monthly Notices of the Royal Astronomical Society] {10.1093/mnras/stad607}, 521, 1871

\bibitem[\protect\citeauthoryear{Baycroft, Triaud  \& Kervella}{Baycroft et~al.}{2023b}]{baycroft_new_2023}
Baycroft T.~A.,  Triaud A. H. M.~J.,   Kervella P.,  2023b, \mn@doi [Monthly Notices of the Royal Astronomical Society] {10.1093/mnras/stad2794}, 526, 2241

\bibitem[\protect\citeauthoryear{Benedict \& Harrison}{Benedict \& Harrison}{2017}]{benedict_hd_2017}
Benedict G.~F.,  Harrison T.~E.,  2017, \mn@doi [The Astronomical Journal] {10.3847/1538-3881/aa6d59}, 153, 258

\bibitem[\protect\citeauthoryear{Bennett et~al.,}{Bennett et~al.}{2016}]{bennett_first_2016}
Bennett D.~P.,  et~al., 2016, \mn@doi [The Astronomical Journal] {10.3847/0004-6256/152/5/125}, 152, 125

\bibitem[\protect\citeauthoryear{Blanco-Cuaresma}{Blanco-Cuaresma}{2019}]{blanco-cuaresma_modern_2019}
Blanco-Cuaresma S.,  2019, \mn@doi [Monthly Notices of the Royal Astronomical Society] {10.1093/mnras/stz549}, 486, 2075

\bibitem[\protect\citeauthoryear{Bouchy, Isambert, Lovis, Boisse, Figueira, Hébrard  \& Pepe}{Bouchy et~al.}{2009a}]{bouchy_charge_2009}
Bouchy F.,  Isambert J.,  Lovis C.,  Boisse I.,  Figueira P.,  Hébrard G.,   Pepe F.,  2009a, \mn@doi [EAS Publications Series] {10.1051/eas/0937031}, 37, 247

\bibitem[\protect\citeauthoryear{Bouchy et~al.,}{Bouchy et~al.}{2009b}]{bouchy_sophie_2009}
Bouchy F.,  et~al., 2009b, \mn@doi [Astronomy and Astrophysics] {10.1051/0004-6361/200912427}, 505, 853

\bibitem[\protect\citeauthoryear{Brewer \& Foreman-Mackey}{Brewer \& Foreman-Mackey}{2018}]{brewer_dnest4_2018}
Brewer B.~J.,  Foreman-Mackey D.,  2018, \mn@doi [Journal of Statistical Software] {10.18637/jss.v086.i07}, 86, 1

\bibitem[\protect\citeauthoryear{Coleman, Nelson  \& Triaud}{Coleman et~al.}{2023}]{coleman_global_2023}
Coleman G. A.~L.,  Nelson R.~P.,   Triaud A. H. M.~J.,  2023, \mn@doi [Monthly Notices of the Royal Astronomical Society] {10.1093/mnras/stad833}, 522, 4352

\bibitem[\protect\citeauthoryear{Coleman, Nelson, Triaud  \& Standing}{Coleman et~al.}{2024}]{coleman_constraining_2024}
Coleman G. A.~L.,  Nelson R.~P.,  Triaud A. H. M.~J.,   Standing M.~R.,  2024, \mn@doi [Monthly Notices of the Royal Astronomical Society] {10.1093/mnras/stad3216}, 527, 414

\bibitem[\protect\citeauthoryear{Collins et~al.,}{Collins et~al.}{2018}]{collins_kelt_2018}
Collins K.~A.,  et~al., 2018, \mn@doi [The Astronomical Journal] {10.3847/1538-3881/aae582}, 156, 234

\bibitem[\protect\citeauthoryear{Correia, Udry, Mayor, Laskar, Naef, Pepe, Queloz  \& Santos}{Correia et~al.}{2005}]{correia_coralie_2005}
Correia A. C.~M.,  Udry S.,  Mayor M.,  Laskar J.,  Naef D.,  Pepe F.,  Queloz D.,   Santos N.~C.,  2005, \mn@doi [Astronomy and Astrophysics] {10.1051/0004-6361:20042376}, 440, 751

\bibitem[\protect\citeauthoryear{Correia, Boué  \& Laskar}{Correia et~al.}{2016}]{correia_secular_2016}
Correia A. C.~M.,  Boué G.,   Laskar J.,  2016, \mn@doi [Celestial Mechanics and Dynamical Astronomy] {10.1007/s10569-016-9709-9}, 126, 189

\bibitem[\protect\citeauthoryear{Courcol et~al.,}{Courcol et~al.}{2015}]{courcol_sophie_2015}
Courcol B.,  et~al., 2015, \mn@doi [Astronomy and Astrophysics] {10.1051/0004-6361/201526329}, 581, A38

\bibitem[\protect\citeauthoryear{Davis et~al.,}{Davis et~al.}{2024}]{davis_eblm_2024}
Davis Y.~T.,  et~al., 2024, \mn@doi [Monthly Notices of the Royal Astronomical Society] {10.1093/mnras/stae842}, 530, 2565

\bibitem[\protect\citeauthoryear{Doolin \& Blundell}{Doolin \& Blundell}{2011}]{doolin_dynamics_2011}
Doolin S.,  Blundell K.~M.,  2011, \mn@doi [Monthly Notices of the Royal Astronomical Society] {10.1111/j.1365-2966.2011.19657.x}, 418, 2656

\bibitem[\protect\citeauthoryear{Doyle et~al.,}{Doyle et~al.}{2011}]{doyle_kepler-16_2011}
Doyle L.~R.,  et~al., 2011, \mn@doi [Science] {10.1126/science.1210923}, 333, 1602

\bibitem[\protect\citeauthoryear{Faria, Santos, Figueira  \& Brewer}{Faria et~al.}{2018}]{faria_kima_2018}
Faria J.~P.,  Santos N.~C.,  Figueira P.,   Brewer B.~J.,  2018, \mn@doi [The Journal of Open Source Software] {10.21105/joss.00487}, 3, 487

\bibitem[\protect\citeauthoryear{Freckelton et~al.,}{Freckelton et~al.}{2024}]{freckelton_bebop_2024}
Freckelton A.~V.,  et~al., 2024, \mn@doi [Monthly Notices of the Royal Astronomical Society] {10.1093/mnras/stae1405}, 531, 4085

\bibitem[\protect\citeauthoryear{Goldberg, Fabrycky, Martin, Albrecht, Deeg  \& Nowak}{Goldberg et~al.}{2023}]{goldberg_5mjup_2023}
Goldberg M.,  Fabrycky D.,  Martin D.~V.,  Albrecht S.,  Deeg H.~J.,   Nowak G.,  2023, \mn@doi [Monthly Notices of the Royal Astronomical Society] {10.1093/mnras/stad2568}, 525, 4628

\bibitem[\protect\citeauthoryear{Gray \& Corbally}{Gray \& Corbally}{1994}]{gray_calibration_1994}
Gray R.~O.,  Corbally C.~J.,  1994, \mn@doi [The Astronomical Journal] {10.1086/116893}, 107, 742

\bibitem[\protect\citeauthoryear{Günther \& Daylan}{Günther \& Daylan}{2019}]{gunther_allesfitter_2019}
Günther M.~N.,  Daylan T.,  2019, Astrophysics Source Code Library, p. ascl:1903.003

\bibitem[\protect\citeauthoryear{Günther \& Daylan}{Günther \& Daylan}{2021}]{gunther_allesfitter_2021}
Günther M.~N.,  Daylan T.,  2021, \mn@doi [The Astrophysical Journal Supplement Series] {10.3847/1538-4365/abe70e}, 254, 13

\bibitem[\protect\citeauthoryear{Hara, Boué, Laskar, Delisle  \& Unger}{Hara et~al.}{2019}]{hara_bias_2019}
Hara N.~C.,  Boué G.,  Laskar J.,  Delisle J.-B.,   Unger N.,  2019, \mn@doi [Monthly Notices of the Royal Astronomical Society, Volume 489, Issue 1, p.738-762] {10.1093/mnras/stz1849}, 489, 738

\bibitem[\protect\citeauthoryear{Hara, Unger, Delisle, Díaz  \& Ségransan}{Hara et~al.}{2022}]{hara_detecting_2022}
Hara N.~C.,  Unger N.,  Delisle J.-B.,  Díaz R.~F.,   Ségransan D.,  2022, \mn@doi [Astronomy and Astrophysics] {10.1051/0004-6361/202140543}, 663, A14

\bibitem[\protect\citeauthoryear{Hara, de Poyferré, Delisle  \& Hoffmann}{Hara et~al.}{2024}]{hara_continuous_2024}
Hara N.~C.,  de Poyferré T.,  Delisle J.-B.,   Hoffmann M.,  2024, \mn@doi [Annals of Applied Statistics] {10.1214/23-AOAS1810}, 18, 749

\bibitem[\protect\citeauthoryear{Heidari}{Heidari}{2022}]{heidari_overcoming_2022}
Heidari N.,  2022, {PhD} {Thesis}, Universit{\textbackslash}'e C{\textbackslash}{\textasciicircum}ote d'Azur; Shahid Beheshti University (Tehran)

\bibitem[\protect\citeauthoryear{Heidari et~al.,}{Heidari et~al.}{2024}]{heidari_sophie_2024}
Heidari N.,  et~al., 2024, \mn@doi [Astronomy and Astrophysics] {10.1051/0004-6361/202347897}, 681, A55

\bibitem[\protect\citeauthoryear{Holman \& Wiegert}{Holman \& Wiegert}{1999}]{holman_long-term_1999}
Holman M.~J.,  Wiegert P.~A.,  1999, \mn@doi [The Astronomical Journal] {10.1086/300695}, 117, 621

\bibitem[\protect\citeauthoryear{Jenkins et~al.,}{Jenkins et~al.}{2016}]{jenkins_tess_2016}
Jenkins J.~M.,  et~al., 2016, \mn@doi [Proceedings of the SPIE] {10.1117/12.2233418}, 9913, 99133E

\bibitem[\protect\citeauthoryear{Kalman}{Kalman}{1996}]{kalman_singularly_1996}
Kalman D.,  1996, \mn@doi [The College Mathematics Journal] {10.1080/07468342.1996.11973744}, 27, 2

\bibitem[\protect\citeauthoryear{Kipping}{Kipping}{2013}]{kipping_parametrizing_2013}
Kipping D.~M.,  2013, \mn@doi [Monthly Notices of the Royal Astronomical Society] {10.1093/mnrasl/slt075}, 434, L51

\bibitem[\protect\citeauthoryear{Kostov et~al.,}{Kostov et~al.}{2016}]{kostov_kepler-1647b_2016}
Kostov V.~B.,  et~al., 2016, \mn@doi [The Astrophysical Journal] {10.3847/0004-637X/827/1/86}, 827, 86

\bibitem[\protect\citeauthoryear{Kostov et~al.,}{Kostov et~al.}{2020}]{kostov_toi-1338_2020}
Kostov V.~B.,  et~al., 2020, \mn@doi [The Astronomical Journal] {10.3847/1538-3881/ab8a48}, 159, 253

\bibitem[\protect\citeauthoryear{Kostov et~al.,}{Kostov et~al.}{2021}]{kostov_tic_2021}
Kostov V.~B.,  et~al., 2021, \mn@doi [The Astronomical Journal] {10.3847/1538-3881/ac223a}, 162, 234

\bibitem[\protect\citeauthoryear{Kurucz}{Kurucz}{2005}]{kurucz_atlas12_2005}
Kurucz R.~L.,  2005, Memorie della Societa Astronomica Italiana Supplementi, 8, 14

\bibitem[\protect\citeauthoryear{Laskar}{Laskar}{1990}]{laskar_chaotic_1990}
Laskar J.,  1990, \mn@doi [Icarus] {10.1016/0019-1035(90)90084-M}, 88, 266

\bibitem[\protect\citeauthoryear{Laskar}{Laskar}{1993}]{laskar_frequency_1993}
Laskar J.,  1993, \mn@doi [Physica D Nonlinear Phenomena] {10.1016/0167-2789(93)90210-R}, 67, 257

\bibitem[\protect\citeauthoryear{Laskar \& Robutel}{Laskar \& Robutel}{2001}]{laskar_high_2001}
Laskar J.,  Robutel P.,  2001, \mn@doi [Celestial Mechanics and Dynamical Astronomy] {10.1023/A:1012098603882}, 80, 39

\bibitem[\protect\citeauthoryear{Li, Holman  \& Tao}{Li et~al.}{2016}]{li_uncovering_2016}
Li G.,  Holman M.~J.,   Tao M.,  2016, \mn@doi [The Astrophysical Journal] {10.3847/0004-637X/831/1/96}, 831, 96

\bibitem[\protect\citeauthoryear{{Lightkurve Collaboration} et~al.,}{{Lightkurve Collaboration} et~al.}{2018}]{lightkurve_collaboration_lightkurve_2018}
{Lightkurve Collaboration} et~al., 2018, Astrophysics Source Code Library, p. ascl:1812.013

\bibitem[\protect\citeauthoryear{Martin}{Martin}{2017}]{martin_circumbinary_2017}
Martin D.~V.,  2017, \mn@doi [Monthly Notices of the Royal Astronomical Society] {10.1093/mnras/stw2851}, 465, 3235

\bibitem[\protect\citeauthoryear{Martin \& Triaud}{Martin \& Triaud}{2014}]{martin_planets_2014}
Martin D.~V.,  Triaud A. H. M.~J.,  2014, \mn@doi [Astronomy \& Astrophysics] {10.1051/0004-6361/201323112}, 570, A91

\bibitem[\protect\citeauthoryear{Martin \& Triaud}{Martin \& Triaud}{2015}]{martin_circumbinary_2015}
Martin D.~V.,  Triaud A. H. M.~J.,  2015, \mn@doi [Monthly Notices of the Royal Astronomical Society] {10.1093/mnras/stv121}, 449, 781

\bibitem[\protect\citeauthoryear{Martin, Mazeh  \& Fabrycky}{Martin et~al.}{2015}]{martin_no_2015}
Martin D.~V.,  Mazeh T.,   Fabrycky D.~C.,  2015, \mn@doi [Monthly Notices of the Royal Astronomical Society] {10.1093/mnras/stv1870}, 453, 3554

\bibitem[\protect\citeauthoryear{Martin et~al.,}{Martin et~al.}{2019}]{martin_bebop_2019}
Martin D.~V.,  et~al., 2019, \mn@doi [Astronomy and Astrophysics] {10.1051/0004-6361/201833669}, 624, A68

\bibitem[\protect\citeauthoryear{Maxted}{Maxted}{2016}]{maxted_ellc_2016}
Maxted P. F.~L.,  2016, \mn@doi [Astronomy and Astrophysics] {10.1051/0004-6361/201628579}, 591, A111

\bibitem[\protect\citeauthoryear{Maxted, Triaud  \& Martin}{Maxted et~al.}{2023}]{maxted_eblm_2023}
Maxted P. F.~L.,  Triaud A. H. M.~J.,   Martin D.~V.,  2023, \mn@doi [Universe] {10.3390/universe9120498}, 9, 498

\bibitem[\protect\citeauthoryear{Meschiari}{Meschiari}{2012}]{meschiari_planet_2012}
Meschiari S.,  2012, \mn@doi [The Astrophysical Journal] {10.1088/2041-8205/761/1/L7}, 761, L7

\bibitem[\protect\citeauthoryear{Moriwaki \& Nakagawa}{Moriwaki \& Nakagawa}{2004}]{moriwaki_planetesimal_2004}
Moriwaki K.,  Nakagawa Y.,  2004, \mn@doi [The Astrophysical Journal] {10.1086/421342}, 609, 1065

\bibitem[\protect\citeauthoryear{Muñoz \& Lai}{Muñoz \& Lai}{2015}]{munoz_survival_2015}
Muñoz D.~J.,  Lai D.,  2015, \mn@doi [Proceedings of the National Academy of Science] {10.1073/pnas.1505671112}, 112, 9264

\bibitem[\protect\citeauthoryear{Orosz et~al.,}{Orosz et~al.}{2012}]{orosz_kepler-47_2012}
Orosz J.~A.,  et~al., 2012, \mn@doi [Science] {10.1126/science.1228380}, 337, 1511

\bibitem[\protect\citeauthoryear{Orosz et~al.,}{Orosz et~al.}{2019}]{orosz_discovery_2019}
Orosz J.~A.,  et~al., 2019, \mn@doi [The Astronomical Journal] {10.3847/1538-3881/ab0ca0}, 157, 174

\bibitem[\protect\citeauthoryear{Paardekooper, Leinhardt, Thébault  \& Baruteau}{Paardekooper et~al.}{2012}]{paardekooper_how_2012}
Paardekooper S.-J.,  Leinhardt Z.~M.,  Thébault P.,   Baruteau C.,  2012, \mn@doi [The Astrophysical Journal] {10.1088/2041-8205/754/1/L16}, 754, L16

\bibitem[\protect\citeauthoryear{Penzlin, Kley  \& Nelson}{Penzlin et~al.}{2021}]{penzlin_parking_2021}
Penzlin A. B.~T.,  Kley W.,   Nelson R.~P.,  2021, \mn@doi [Astronomy and Astrophysics] {10.1051/0004-6361/202039319}, 645, A68

\bibitem[\protect\citeauthoryear{Pepe, Mayor, Galland, Naef, Queloz, Santos, Udry  \& Burnet}{Pepe et~al.}{2002}]{pepe_coralie_2002}
Pepe F.,  Mayor M.,  Galland F.,  Naef D.,  Queloz D.,  Santos N.~C.,  Udry S.,   Burnet M.,  2002, \mn@doi [Astronomy and Astrophysics] {10.1051/0004-6361:20020433}, 388, 632

\bibitem[\protect\citeauthoryear{Pepper et~al.,}{Pepper et~al.}{2007}]{pepper_kilodegree_2007}
Pepper J.,  et~al., 2007, \mn@doi [Publications of the Astronomical Society of the Pacific] {10.1086/521836}, 119, 923

\bibitem[\protect\citeauthoryear{Perruchot et~al.,}{Perruchot et~al.}{2008}]{perruchot_sophie_2008}
Perruchot S.,  et~al., 2008, \mn@doi [Proceedings of the SPIE] {10.1117/12.787379}, 7014, 70140J

\bibitem[\protect\citeauthoryear{Pierens \& Nelson}{Pierens \& Nelson}{2008}]{pierens_formation_2008}
Pierens A.,  Nelson R.~P.,  2008, \mn@doi [Astronomy and Astrophysics] {10.1051/0004-6361:200809453}, 483, 633

\bibitem[\protect\citeauthoryear{Pierens, Nelson  \& McNally}{Pierens et~al.}{2021}]{pierens_vertical_2021}
Pierens A.,  Nelson R.~P.,   McNally C.~P.,  2021, \mn@doi [Monthly Notices of the Royal Astronomical Society] {10.1093/mnras/stab2853}, 508, 4806

\bibitem[\protect\citeauthoryear{Pollacco et~al.,}{Pollacco et~al.}{2006}]{pollacco_wasp_2006}
Pollacco D.~L.,  et~al., 2006, \mn@doi [Publications of the Astronomical Society of the Pacific] {10.1086/508556}, 118, 1407

\bibitem[\protect\citeauthoryear{Pollacco et~al.,}{Pollacco et~al.}{2008}]{pollacco_wasp-3b_2008}
Pollacco D.,  et~al., 2008, \mn@doi [Monthly Notices of the Royal Astronomical Society] {10.1111/j.1365-2966.2008.12939.x}, 385, 1576

\bibitem[\protect\citeauthoryear{Qian et~al.,}{Qian et~al.}{2011}]{qian_detection_2011}
Qian S.~B.,  et~al., 2011, \mn@doi [Monthly Notices of the Royal Astronomical Society] {10.1111/j.1745-3933.2011.01045.x}, 414, L16

\bibitem[\protect\citeauthoryear{Ricker et~al.,}{Ricker et~al.}{2015}]{ricker_transiting_2015}
Ricker G.~R.,  et~al., 2015, \mn@doi [Journal of Astronomical Telescopes, Instruments, and Systems] {10.1117/1.JATIS.1.1.014003}, 1, 014003

\bibitem[\protect\citeauthoryear{Roy \& Vetterli}{Roy \& Vetterli}{2007}]{roy_effective_2007}
Roy O.,  Vetterli M.,  2007, 15th European Signal Processing Conference, pp 606--610

\bibitem[\protect\citeauthoryear{Sairam et~al.,}{Sairam et~al.}{2024}]{sairam_new_2024}
Sairam L.,  et~al., 2024, \mn@doi [Monthly Notices of the Royal Astronomical Society] {10.1093/mnras/stad3136}, 527, 2261

\bibitem[\protect\citeauthoryear{Sbordone, Bonifacio, Castelli  \& Kurucz}{Sbordone et~al.}{2004}]{sbordone_atlas_2004}
Sbordone L.,  Bonifacio P.,  Castelli F.,   Kurucz R.~L.,  2004, \mn@doi [Memorie della Societa Astronomica Italiana Supplementi] {10.48550/arXiv.astro-ph/0406268}, 5, 93

\bibitem[\protect\citeauthoryear{Sebastian, Triaud  \& Brogi}{Sebastian et~al.}{2024a}]{sebastian_saltire_2024}
Sebastian D.,  Triaud A. H. M.~J.,   Brogi M.,  2024a, \mn@doi [Monthly Notices of the Royal Astronomical Society] {10.1093/mnras/stad3765}, 527, 10921

\bibitem[\protect\citeauthoryear{Sebastian et~al.,}{Sebastian et~al.}{2024b}]{sebastian_eblm_2024}
Sebastian D.,  et~al., 2024b, \mn@doi [Monthly Notices of the Royal Astronomical Society] {10.1093/mnras/stae459}, 530, 2572

\bibitem[\protect\citeauthoryear{Sebastian et~al.,}{Sebastian et~al.}{2025}]{sebastian_eblm_2025}
Sebastian D.,  et~al., 2025, \mn@doi [Monthly Notices of the Royal Astronomical Society] {10.1093/mnras/staf863}, 540, 2914

\bibitem[\protect\citeauthoryear{Snellen, de Kok, de Mooij  \& Albrecht}{Snellen et~al.}{2010}]{snellen_orbital_2010}
Snellen I. A.~G.,  de Kok R.~J.,  de Mooij E. J.~W.,   Albrecht S.,  2010, \mn@doi [Nature] {10.1038/nature09111}, 465, 1049

\bibitem[\protect\citeauthoryear{Speagle}{Speagle}{2020}]{speagle_dynesty_2020}
Speagle J.~S.,  2020, \mn@doi [Monthly Notices of the Royal Astronomical Society] {10.1093/mnras/staa278}, 493, 3132

\bibitem[\protect\citeauthoryear{Standing et~al.,}{Standing et~al.}{2022}]{standing_bebop_2022}
Standing M.~R.,  et~al., 2022, \mn@doi [Monthly Notices of the Royal Astronomical Society] {10.1093/mnras/stac113}, 511, 3571

\bibitem[\protect\citeauthoryear{Standing et~al.,}{Standing et~al.}{2023}]{standing_radial-velocity_2023}
Standing M.~R.,  et~al., 2023, \mn@doi [Nature Astronomy] {10.1038/s41550-023-01948-4}, 7, 702

\bibitem[\protect\citeauthoryear{Triaud}{Triaud}{2011}]{triaud_constraints_2011}
Triaud A. H. M.~J.,  2011, Ph.{D}. thesis, \url {https://ui.adsabs.harvard.edu/abs/2011PhDT........22T}

\bibitem[\protect\citeauthoryear{Triaud et~al.,}{Triaud et~al.}{2013}]{triaud_eblm_2013}
Triaud A. H. M.~J.,  et~al., 2013, \mn@doi [Astronomy and Astrophysics] {10.1051/0004-6361/201219643}, 549, A18

\bibitem[\protect\citeauthoryear{Triaud et~al.,}{Triaud et~al.}{2017}]{triaud_eblm_2017}
Triaud A. H. M.~J.,  et~al., 2017, \mn@doi [Astronomy and Astrophysics] {10.1051/0004-6361/201730993}, 608, A129

\bibitem[\protect\citeauthoryear{Triaud et~al.,}{Triaud et~al.}{2022}]{triaud_bebop_2022}
Triaud A. H. M.~J.,  et~al., 2022, \mn@doi [Monthly Notices of the Royal Astronomical Society] {10.1093/mnras/stab3712}, 511, 3561

\bibitem[\protect\citeauthoryear{Trotta}{Trotta}{2008}]{trotta_bayes_2008}
Trotta R.,  2008, \mn@doi [Contemporary Physics] {10.1080/00107510802066753}, 49, 71

\bibitem[\protect\citeauthoryear{Wang \& Liu}{Wang \& Liu}{2024}]{wang_photo-dynamical_2024}
Wang M.-T.,  Liu H.-G.,  2024, \mn@doi [The Astronomical Journal] {10.3847/1538-3881/ad4a60}, 168, 31

\bibitem[\protect\citeauthoryear{Welsh et~al.,}{Welsh et~al.}{2012}]{welsh_transiting_2012}
Welsh W.~F.,  et~al., 2012, \mn@doi [Nature] {10.1038/nature10768}, 481, 475

\bibitem[\protect\citeauthoryear{Winn \& Fabrycky}{Winn \& Fabrycky}{2015}]{winn_occurrence_2015}
Winn J.~N.,  Fabrycky D.~C.,  2015, \mn@doi [Annual Review of Astronomy and Astrophysics, vol. 53, p.409-447] {10.1146/annurev-astro-082214-122246}, 53, 409

\makeatother
\end{thebibliography}




\appendix

\section{Appendix}

\begin{table}
    \centering
    \caption{Radial velocity observations taken with SOPHIE between 2018-11-13 and 2024-11-18. Radial velocity outliers are indicated with one of two flags X: those that are vast outliers found in the preliminary run and removed, these are likely wrong pointings of the telescope, and Y: those that are not excluded from the fit but accounted for with the Student's t and have a ratio of \(L_T/L_N >= 10\) in the second analysis. Spectra excluded from the HRCCS analysis are given the flag A (wrong pointing), B (low SNR), or C (blended phases).}
    \begin{tabular}{c|c|c|c}
        Flag & Time & \(V_{\rm rad}\) & \(\sigma_{V_{\rm rad}}\) \\
         & [BJD - 2\,400\,000] & [\(km\,s^{-1}\)] & [\(km\,s^{-1}\)] \\
        \hline
        & 58436.63359 & 43.7735 & 0.0039 \\
        & 58437.68801 & 42.9949 & 0.0018 \\
        & 58536.46316 & 5.4022 & 0.0024 \\
        & 58539.46867 & 21.9657 & 0.0021 \\
        & 58541.46505 & 40.2170 & 0.0018 \\
        & 58556.34359 & 43.7120 & 0.0021 \\
        & 58560.44450 & 16.9170 & 0.0029 \\
        & 58564.34291 & 8.8614 & 0.0018 \\
        & 58794.64166 & 42.7380 & 0.0026 \\
        & 58804.69818 & 30.9787 & 0.0024 \\
        & 58828.65843 & 8.6339 & 0.0025 \\
        Y& 58855.59869 & 11.8996 & 0.0028 \\
        & 58894.46828 & 7.3007 & 0.0020 \\
        & 58899.35178 & 43.9156 & 0.0026 \\
        C& 58903.33834 & 22.9174 & 0.0020 \\
        & 58907.45466 & 6.4592 & 0.0027 \\
        B& 58912.32987 & 43.4670 & 0.0129 \\
        & 58917.38073 & 16.4726 & 0.0031 \\
        & 58922.32180 & 17.3861 & 0.0028 \\
        X,A& 59130.68250 & -3.6578 & 0.0361 \\
        & 59155.65324 & 13.9372 & 0.0018 \\
        ... & & & \\
        \hline
    \end{tabular}
    \label{tab:rv_data}
\end{table}

\begin{table}
    \centering
    \caption{Prior distributions used in the {\tt kima} analysis. \(\pazocal{N}\) is a Normal/Gaussian distribution, \(\pazocal{U}\) is a Uniform distribution, \(\pazocal{LU}\) is a LogUniform distribution, \(\pazocal{MLU}\) is a ModifiedLogUniform distribution (LogUniform between the two values and Uniform between 0 and the lower value), \(\pazocal{K}\) is a Kumaraswamy distribution (a computationally efficient approximation to a \(\beta\) distribution).}
    \begin{tabular}{c|c|c}
        Parameter & Units & Prior distribution \\
        \hline
        \(P_{\rm bin}\) & [days] & \(\pazocal{N}(13.2177,0.001)\) \\
        \(K_{\rm bin}\) & [\({\rm km\,s^{-1}}\)] & \(\pazocal{N}(19.366,0.05)\) \\
        \(e_{\rm bin}\) & & \(\pazocal{N}(0.063,0.01)\) \\
        \(\omega_{\rm bin}\) & [rad] & \(\pazocal{N}(4.9,0.1)\) \\
        \(M_{0, {\rm bin}}\) & [rad] & \(\pazocal{U}(0,2\pi)\) \\
        \(\dot{\omega}_{\rm bin}\) & [\({\rm arcsec\,yr^{-1}}\)] & \(\pazocal{N}(0,1000)\) \\
        \hline
        \(P_{\rm pl}\) & [days] & \(\pazocal{LU}(50,5490)\) \\
        \(K_{\rm pl}\) & [\({\rm m\,s^{-1}}\)] & \(\pazocal{MLU}(0.1,1000)\) \\
        \(e_{\rm pl}\) & & \(\pazocal{K}(0.867,3.03)\) \\
        \(\omega_{\rm pl}\) & [rad] & \(\pazocal{U}(0,2\pi)\) \\
        \(M_{0, {\rm pl}}\) & [rad] & \(\pazocal{U}(0,2\pi)\) \\
        \hline
        \(V_{\rm sys}\) & [\({\rm km\,s^{-1}}\)] & \(\pazocal{U}(23.968,24.968)\) \\
        Jitter & [\({\rm m\,s^{-1}}\)] & \(\pazocal{MLU}(0.1,100)\) \\
        \(\nu\) & & \(\pazocal{LU}(2,1000)\) \\
         
    \end{tabular}
    \label{tab:priors}
\end{table}

\begin{figure*}
    \centering
    \includegraphics[width=0.49\linewidth]{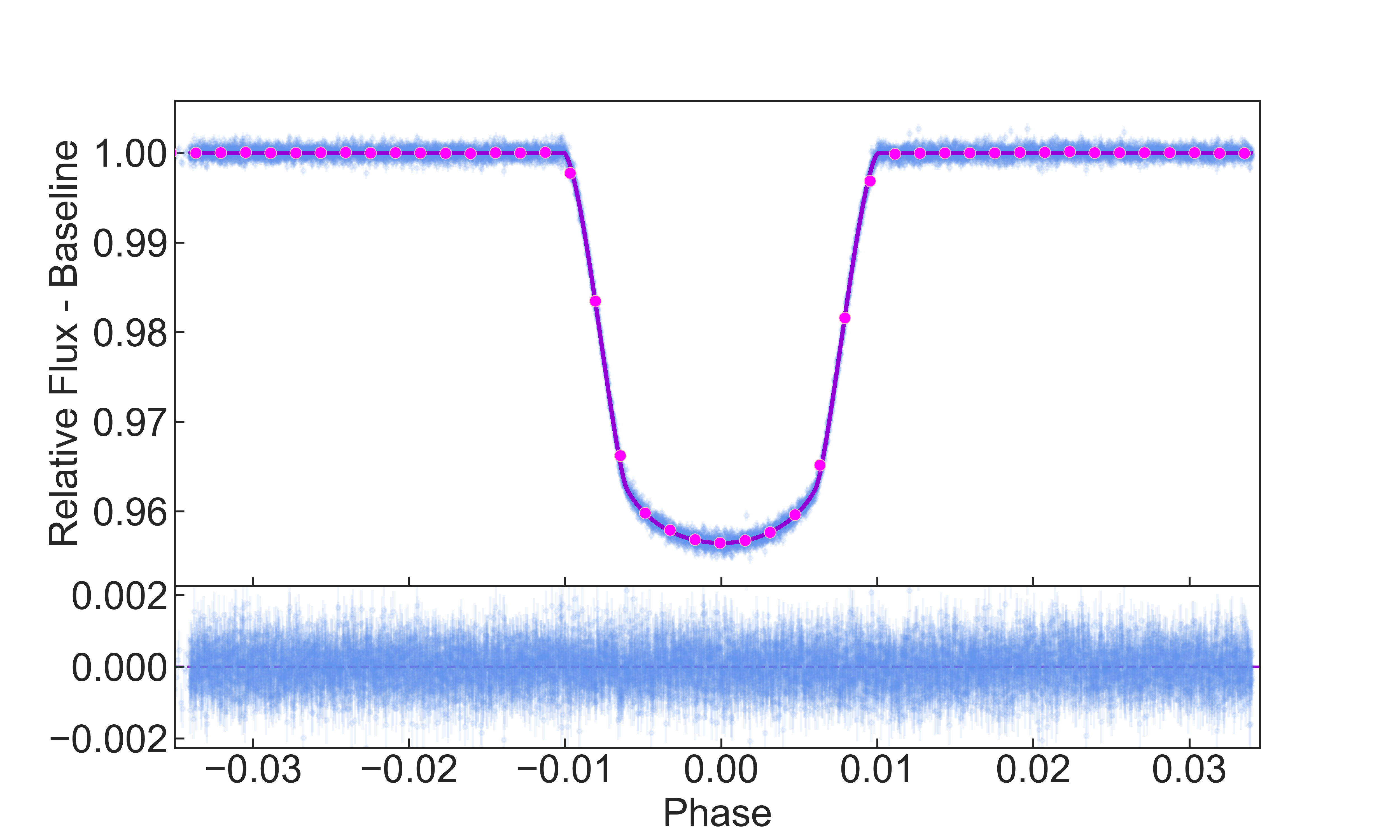}
    \includegraphics[width=0.49\linewidth]{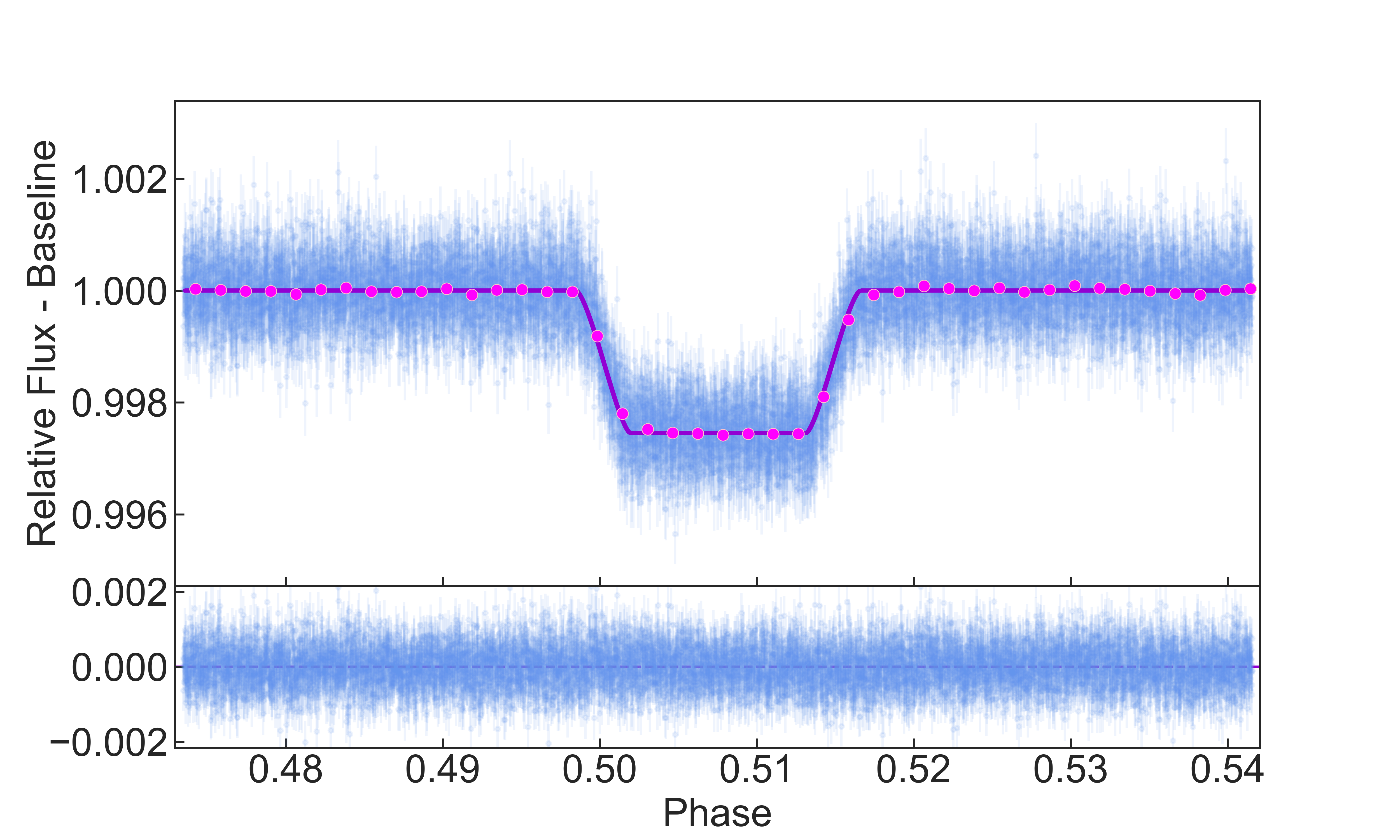}
    \caption{Left panels: Phase folded primary eclipses for the binary and the residuals for the best fit model. Right panels: Phase folded secondary eclipses for the binary and the residuals for the best fit model. In both plots, 50 random posterior samples are displayed.}
    \label{fig:phase_folded_phot}
\end{figure*}

\begin{figure}
    \centering
    \includegraphics[width=0.98\linewidth]{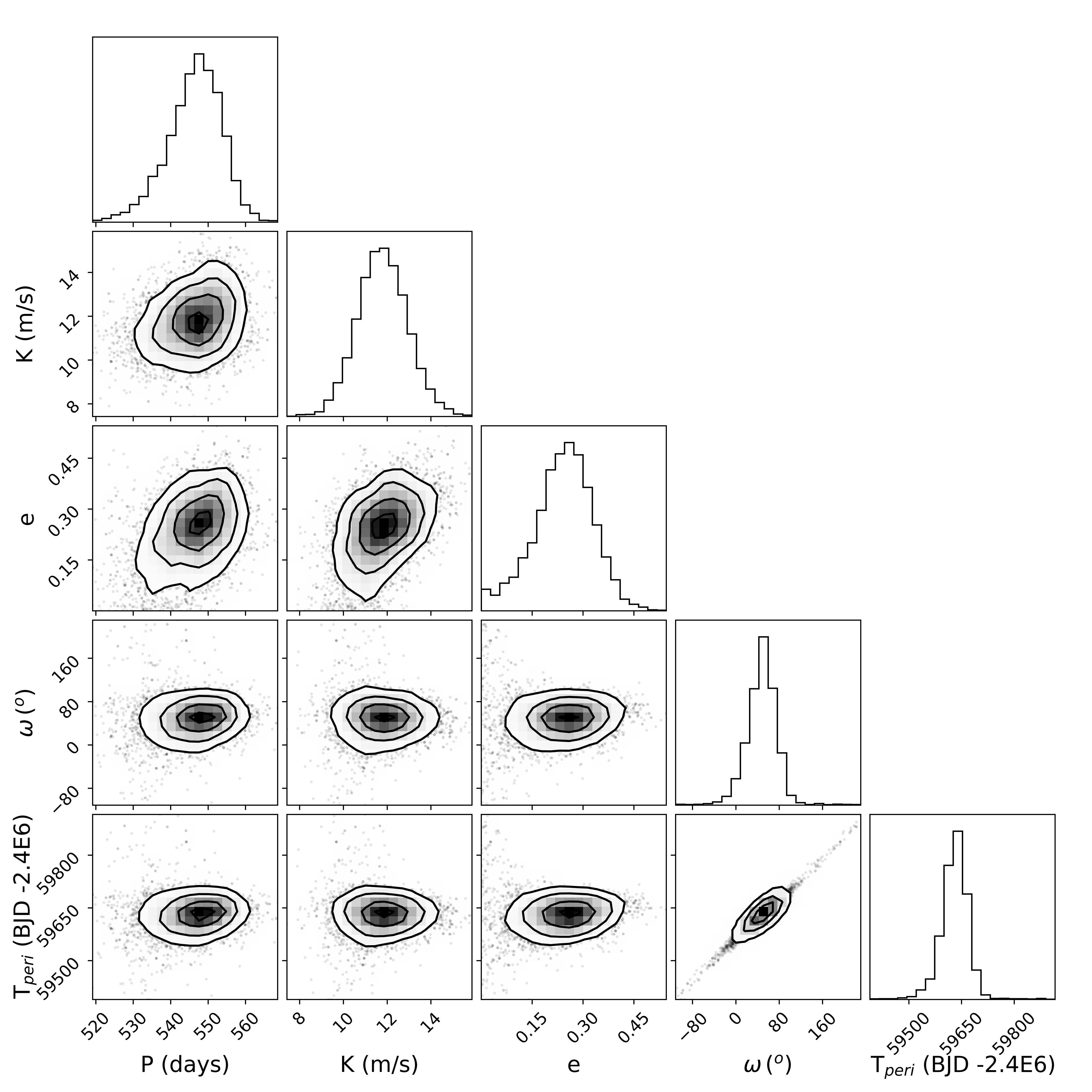}
    \caption{Corner plot showing posterior distributions and correlations for the 1-planet model.}
    \label{fig:corner_1pl}
\end{figure}

\begin{figure}
    \centering
    \includegraphics[width=0.98\linewidth]{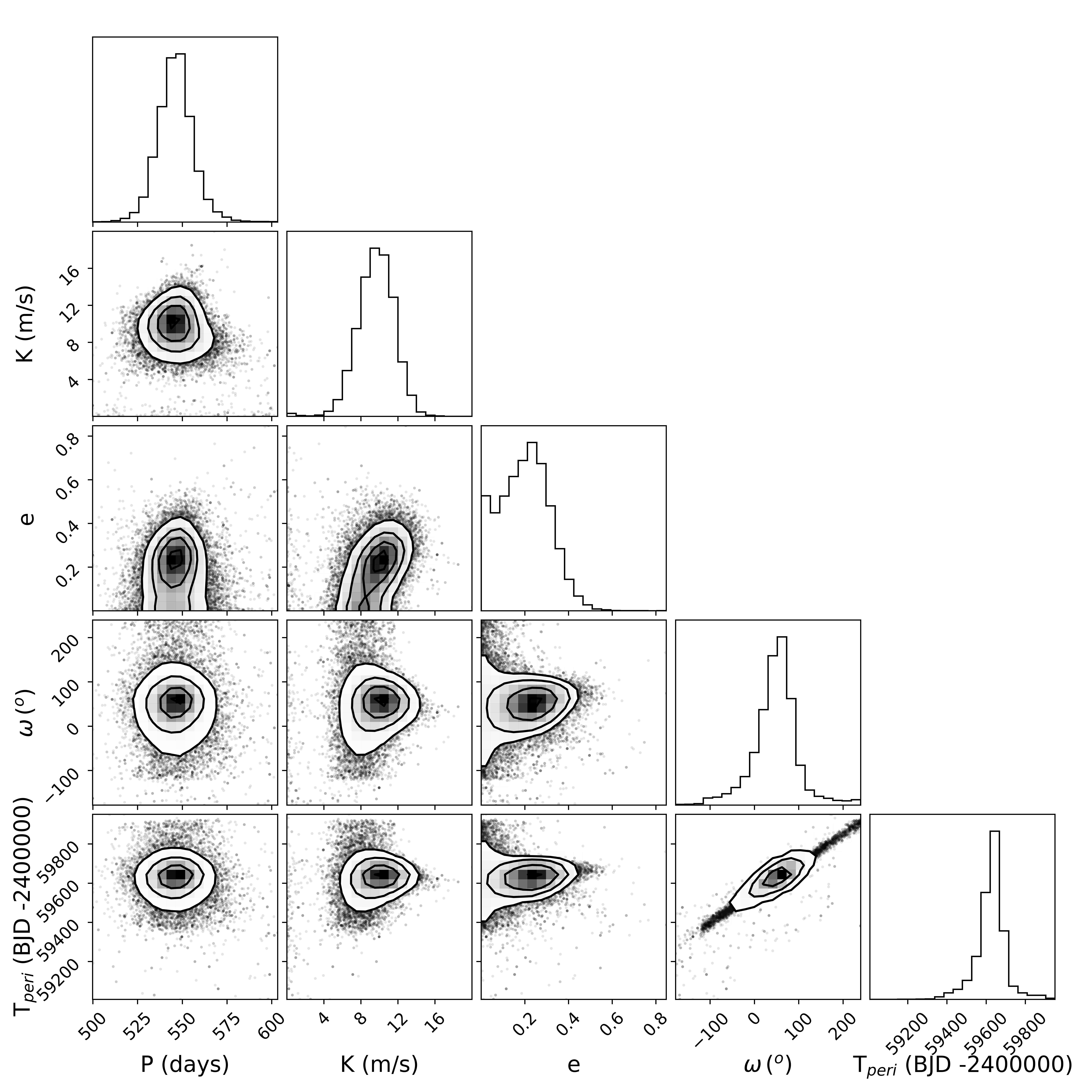}
    \caption{Corner plot showing posterior distributions and correlations for planet 1 in the 2-planet model.}
    \label{fig:corner_2pl_1}
\end{figure}

\begin{figure}
    \centering
    \includegraphics[width=0.98\linewidth]{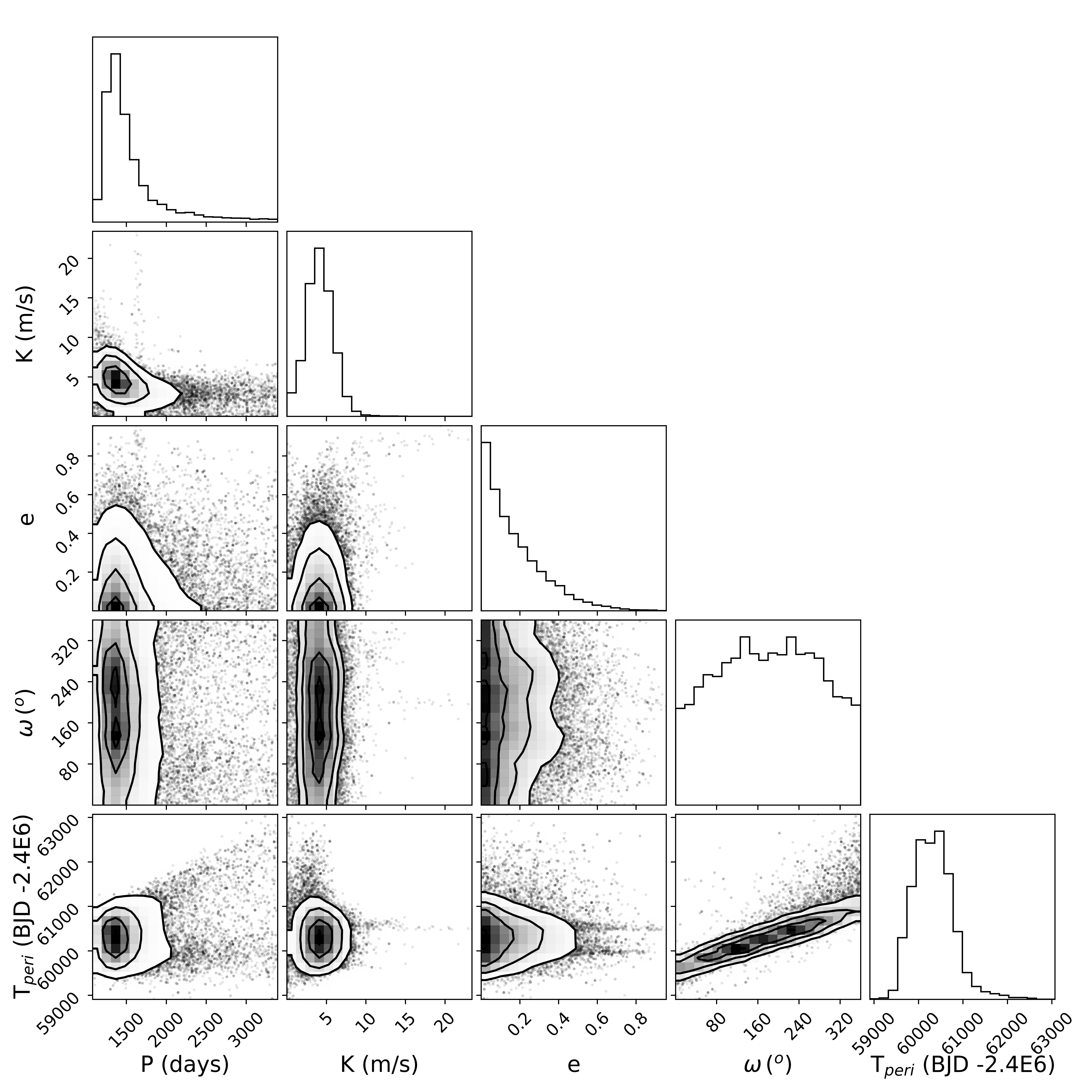}
    \caption{Corner plot showing posterior distributions and correlations for planet 2 in the 2-planet model.}
    \label{fig:corner_2pl_2}
\end{figure}

\begin{table}
    {\centering
    \caption{Orbital parameters of BEBOP-3 b, marginalised over the candidate and undetected potential planets by inlcuding the posterior samples from \(N_P = 1\), \(N_P = 2\), and \(N_P = 3\) . Note that these are mean orbital elements rather osculating elements. The reference time (BJD) used in the analysis is 2459766.975506.}
    \begin{tabular}{l|c|c|c}
        Parameter & Units & Value & Note \\
        \hline
        \(P\) & [days] & \(546.0^{+8.7}_{-9.2}\) & \\
        \(K\) & [\({\rm m\,s^{-1}}\)] & \(9.8^{+1.8}_{-2.0}\) & \\
        \(e\) & & \(0.20^{+0.11}_{-0.13}\) & \\
        \(\omega\) & [rad] & \(0.90^{+0.59}_{-0.74}\) & \({(b)}\) \\
        \(\lambda_0\) & [rad] & \(2.79\pm0.26\) & \({(b)}\) \\
        \({T_{\rm peri}}\) & [BJD-2450000] & \(59633^{+45}_{-62}\) & \\
        \(M\) & [\({\rm M_{Jup}}\)] & \(0.469^{+0.083}_{-0.090}\) & \({(a)}\) \\
        \(a\) & [AU] & \(1.443\pm0.019\) & \\
        \hline
    \end{tabular}\\}
    \(^{(a)}\) True mass of the planet under the assumption it is exactly coplanar with the binary.\\ \(^{(b)}\) the argument of pericentre \(\omega\) and the true longitude at the reference time \(\lambda_0\) are those of the orbit of the binary around the centre-of mass of the planet-binary two body orbit, to convert to the parameters for the planet, \(\pi\) should be subtracted.
    \label{tab:pl_pars_marg}
\end{table}

\begin{landscape}

\begin{table}
\centering
\caption{The BEBOP catalogue of circumbinary planets with radial velocity measurements. \(1\sigma\) uncertainties on parameters are provided at the level of the last two significant digits. \(^x\) Transit detection from \citet{doyle_kepler-16_2011} .\(^y\) Transit detection and parameters from \citet{kostov_toi-1338_2020} .\(^z\) Transit detection from \citet{kostov_tic_2021}.}
\begin{tabular}{l|c|c|c|c|c|c|c|c|c|r}
     Catalogue entry & Other names & \(P_{\rm bin}\) & \(M_{\rm pri}\) & \(M_{\rm sec}\) & \(e_{\rm bin}\) & \(P_{\rm pl}\) & \(M_{\rm pl}\) & \(e_{\rm pl}\)  & Source of RVs & TIC ID \\
      &  & [days] & [\(M_{\rm \odot}\)] & [\(M_{\rm \odot}\)] &  & [days] & \(M_{\rm Jup}\) &  &  &   \\
     BEBOP-0 b & Kepler-16 b\(^x\) & \(41.077772(51)\) & \(0.654(17)\) & \(0.1964(31)\) & \(0.15994(10)\) & \(226.0(1.7)\) & \(0.313(39)\) & \(\leq0.21\) & \citet{triaud_bebop_2022} & 299096355\\
     BEBOP-1 b & TOI-1338 b\(^y\)  & \(14.6085579(57)\) & \(1.098(17)\) & \(0.307(3)\) & \(0.155522(29)\) & \(95.174(35)^y\) & \(<0.0685(29)\) & \(0.0880(43)^y\) & \citet{standing_radial-velocity_2023} & 260128333\\
     BEBOP-1 c & TOI-1338 c & \(14.6085579(57)\) & \(1.098(17)\) & \(0.307(3)\) & \(0.155522(29)\) & \(215.5(3.3)\) & \(0.205(37)\) & \(\leq0.16\) & \citet{standing_radial-velocity_2023} & 260128333 \\
     BEBOP-2 b & TIC 172900988 b\(^z\) & \(19.657878(34)\) & \(1.23681(39)\) & \(1.20207(33)\) & \(0.448234(90)\) & \(151.2(1.8)\) & \(1.90(25)\) & \(\leq0.11\) & \citet{sairam_new_2024} & 172900988 \\
     BEBOP-3 b & & \(13.2176657(27)\) & \(1.083(26)\) & \(0.3615(39)\) & \(0.063255(54)\) & \(547.0^{(+6.2)}_{(-7.6)}\) & \(0.558^{(+51)}_{(-48)}\) & \(0.247^{(+77)}_{(-89)}\) & This work & 289949453 \\
\end{tabular}
\label{tab:catalogue}
\end{table}
\end{landscape}



\bsp	
\label{lastpage}
\end{document}